\documentclass[pra,showpacs,showkeywords,graphics,epsfigure,superscriptaddresss]{revtex4}

\pdfoutput=1

\usepackage{graphicx}
\usepackage{graphics}
\usepackage{bbm}
\usepackage{amsmath}
\usepackage{amsfonts}
\usepackage{amssymb}
\usepackage{latexsym}

\usepackage{braket}
\usepackage{amsbsy}
\usepackage{amsthm}
\usepackage{bm}
\usepackage{dsfont} 

\newcommand{\beq}{\begin{eqnarray}}
\newcommand{\eeq}{\end{eqnarray}}

\newcommand{\tr}{\textnormal{Tr}}

\newtheorem{theorem}{Theorem}

\newtheorem{corollary}{Corollary}
\newtheorem{definition}{Definition}

\newcommand{\B}{\mathcal{B}}
\newcommand{\Ha}{\mathcal{H}}

\newcommand{\V}{\mathcal{V}}

\begin{document}

\title{John Bell and the Nature of the Quantum World}
\author{Reinhold A. Bertlmann}
\affiliation{University of Vienna, Faculty of Physics, Boltzmanngasse 5, 1090 Vienna, Austria}
\email{Reinhold.Bertlmann@univie.ac.at}

\begin{abstract}

I present my encounter with John Bell at CERN, our collaboration and joint work in particle physics. I also will recall our quantum debates and give my personal view on Bell's fundamental work on quantum theory, in particular, on contextuality and nonlocality of quantum physics. Some mathematical and geometric aspects of entanglement are discussed as influence of Bell's Theorem. Finally, I make some historical comments on the experimental side of Bell inequalities.

\end{abstract}


\maketitle


\section{Prologue}

\subsection{Encounter with John Bell}

At the end of 1977 I received a letter from CERN that the Selection Committee had appointed me to be a Fellow at CERN and that my employment would begin on April 1st 1978, a memorable day. At the end of March, 1978, I moved from Vienna to Geneva with all my scripts and books, heavy luggage at that time. I introduced myself at CERN's Theory Division and began to attend all the activities and events like the Theoretical Seminar. Already in one of the first weeks, after one of these seminars, when drinking a welcome tea in the Common Room, an impressive man with metal-rimmed glasses, red hair, and a beard appeared in the door, looked around, fixed me, and approached straightaway. \emph{``I'm John Bell, who are you?''} he introduced himself. Of course, being a young postdoc, I was quite speechless to have this famous physicist in front of me, renowned for his work in weak interactions and gauge theories. With a bit of stuttering I replied, \emph{``I am Reinhold Bertlmann from Vienna, Austria.''} \emph{``What are you working on?''} was his next question and I answered \emph{``Quarkonium...''}. Bell began to smile gently and we immediately fell into a lively discussion about bound states of quark-antiquark systems, a very popular subject at that time. The discussions continued in front of the blackboard in his office and gave rise to common calculations---a happy and fruitful collaboration and friendship began.

\subsection{Joint Work in Particle Physics}

The 1970s was a glorious time for experimental particle physics, in particular, several spectra of hadronic narrow resonances, like the $J/\psi$ charmonium and $\Upsilon$ bottonium resonances, have been discovered in $e^+ e^-$ collisions. The production and the properties of these particles had to be understood.

The first problem we attacked was how to understand the production of hadrons in $e^+ e^-$ collisions. Hadrons consist of quarks and antiquarks, thus the $e^+ e^-$ collisions actually produce quark-antiquark ($q \bar{q}$) pairs. The idea was that at high energies, which correspond to short distances, the $q \bar{q}$ pairs behave as quasi-free particles that produce a flat cross-section. However, at low energies the quarks propagate larger distances, are confined and generate bound states---called quarkonium---which show up as bumps, resonances, in the cross-section.

We followed an idea that goes back to J.J. Sakurai \cite{Sakurai1973}. We smeared each resonance in energy appropriately so that the cross-section over one resonance agreed already with the correspondingly averaged asymptotic result. We called it \textit{local duality} \cite{BellBertlmannDual1980, Bertlmann-ActaPhysAustr1981} since two seemly different features appear as the \textit{dual} aspects of one and the same reality.

The duality relation we can understand quite easily. Allowing for an energy spread means---via the uncertainty relation---that we focus on short times. But for short times the corresponding wave does not spread far enough to feel the details of the long distances, the confining potential. So this part can be neglected and the wave function at the origin of the bound state, which determines the leptonic width or area of a resonance, matches the averaged quasi-free $q \bar{q}$ pair. However, if we want to push the idea of duality even further in order to become sensitive to the position of the bound state, the resonance, in the mass spectrum then we have to include the contributions of the larger distances into the wave function. That means, we must have some information about confinement.\\

\begin{figure}
\begin{center}
\includegraphics[width=0.5\textwidth]{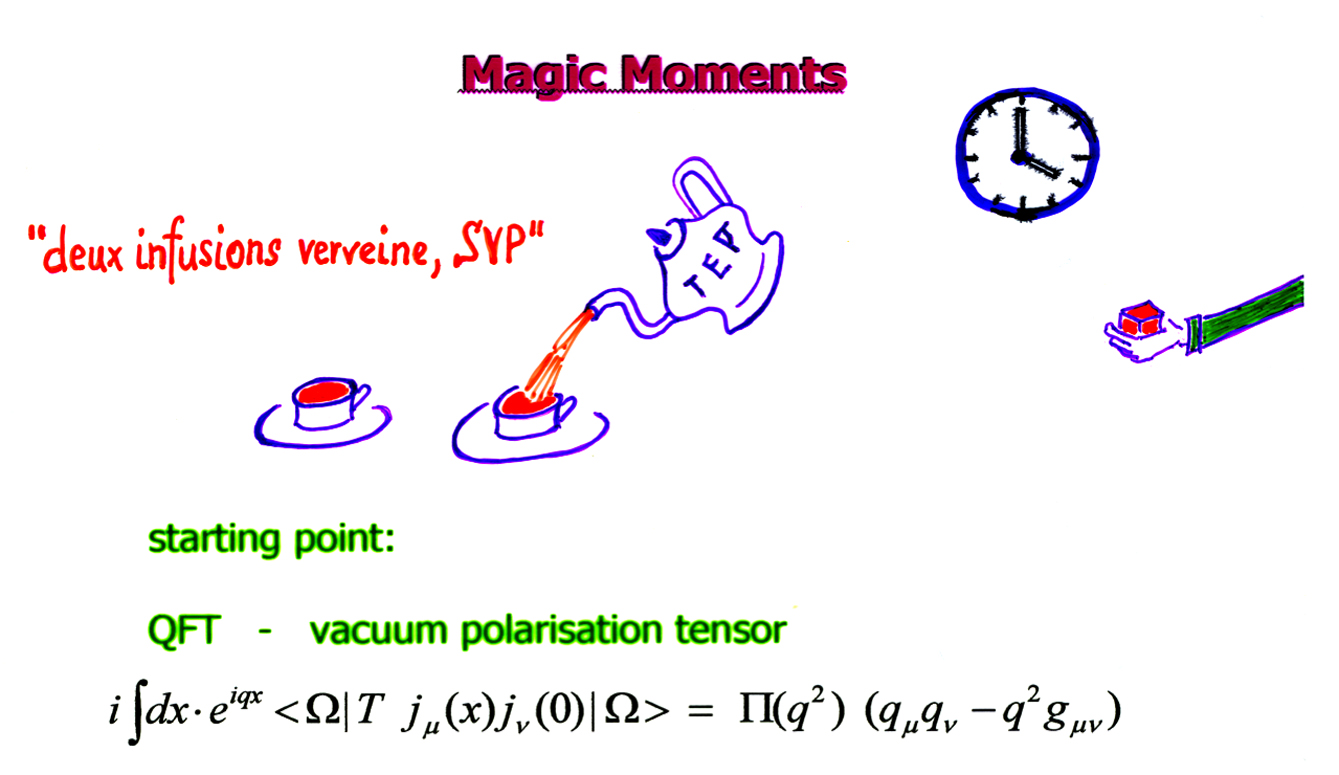}\\
\includegraphics[width=0.5\textwidth]{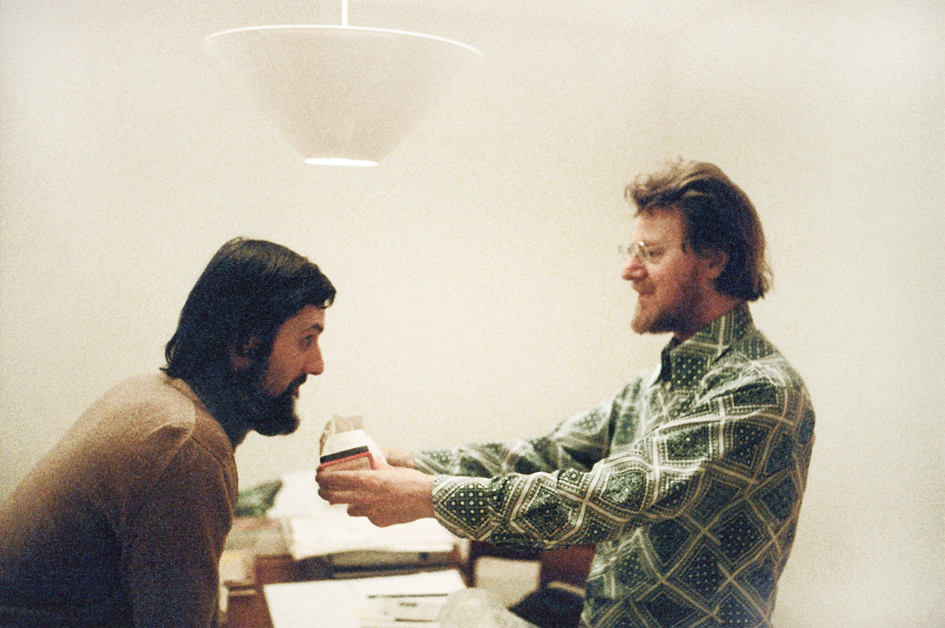}
\caption{Bell and Bertlmann choosing the right sort of tea in 1980. Foto: \copyright Renate Bertlmann.}
	\label{fig:Reinhold+John choosing tea}
\end{center}
\end{figure}

That's the subject Bell and I got interested in next and it resulted in a wonderful collaboration which we called \emph{Magic Moments \/} \cite{BellBertlmannMagic1981}. As a true Irishman John always had to drink a \emph{4~o'clock tea}\,, see Fig.~\ref{fig:Reinhold+John choosing tea}, and this we regularly practiced in the CERN Cafeteria. He ordered with his typical Irish accent, \emph{``deux infusions verveine, s'il vous pl\^ait"}, John's favorite tea. There, in a relaxed atmosphere, we talked not only about physics but also about politics, philosophy, and when Renate Bertlmann joined us we also had heated debates about modern art.\\

How could we include confinement in our duality concept? Our starting point was the vacuum polarization tensor, the vacuum expectation value of the time ordered product of two quark currents in quantum field theory, quantum chromo dynamics. This quantity was proportional to the hadronic cross-section and calculable within perturbation theory. At that time a Russian group including Shifman, Vainshtein, and Zakharov \cite{SVZ1979}, claimed that the so-called gluon condensate, the vacuum expectation value of two gluon fields, is responsible for the influence of confinement. We wanted to examine this idea further.

Approximating quantum field theory by potential theory, we could calculate both the perturbative \emph{and} the exact result. For the energy smearing we chose an exponential as the weight function, which is called a \emph{moment} by the mathematicians. In this case, the procedure corresponds to perturbation theory of a Hamiltonian with respect to an imaginary time. I found this quite fascinating. The ground state level could be extracted by using the logarithmic derivative of a moment, which approaches the ground state energy for large (imaginary) times since the contributions of the higher levels are cut off. In this way we were able to predict the ground states of charmonium (the $J/\psi$ resonances) and of bottonium (the $\Upsilon$ resonances) to a high accuracy, quantitatively within $10 \%$ \cite{BellBertlmannMagic1981, BertlmannCharmonium-PhysLett1981, BertlmannCharmonium-NuclPhys1982}. Thus in this respect we could demonstrate the success of the procedure.\\

John and I were working within potential theory which functioned surprisingly well \cite{BellBertlmann-PhysLett1984}. Therefore it was quite natural for us to ask whether one can attach a potential to the occurrence of the gluon condensate. Indeed, we found ways to do this \cite{BellBertlmann-SVZ1981, BellBertlmann-LV1983}.

One way was to work within the moments, which regularize the divergence of the long-distance part of the gluon propagator, the gluon condensate contribution. In this case a static, nonrelativistic potential containing the gluon condensate can be extracted, which is called in the literature the \emph{equivalent potential of Bell and Bertlmann} \cite{BellBertlmann-SVZ1981}. The short-distance part is the well-known Coulomb potential, whereas the long-distance component, the gluon condensate contribution, emerges as a quartic potential $m r^4$ and is mass- i.e. flavour-dependent. In this respect it differs considerably from the familiar mass-independent, rather flat potential models \cite{QuiggRosner1979, KrammerKrasemann1979, GrosseMartin1980, LuchaSchoeberlGromes1991}. However, for a final comparison with potential models one has to go further and take into account the higher order fluctuations \cite{Bertlmann-GGGG1984}.

By studying very heavy quarkonium systems, which describe e.g. bottonium (the $\Upsilon$ resonances), John and I also found another way to extract a potential from the gluon condensate effect~\cite{BellBertlmann-LV1983}. In that case the low-lying bound states, because of their small size, will be dominated by the Coulomb potential. The condensate effect, an external colour field representing the gluon, can be added as a small perturbation. Leutwyler \cite{Leutwyler1981} and Voloshin \cite{Voloshin1979} had considered such a colour-electric Stark effect and calculated the energy spectrum for all quantum numbers $n$ and $l$. John and myself, on the other hand, were able to construct a gluonic potential, which by perturbing the Coulomb states provides the energy spectrum of Leutwyler-Voloshin. The leading term of this potential, the infinite mass limit, has a cubic $r^3$ dependence and is therefore mass-independent. But for finite masses there are further corrections proportional to $\frac{r^2}{m}, \frac{r}{m^2}, \frac{constant}{m^3}$ such that the potential becomes flavour-dependent again.\\

In conclusion, no adequate bridge was found between a field theory containing the gluon condensate, quantum chromo dynamics, on one side, and popular potential models on the other. For an overview of this field I refer to Ref.~\cite{BertlmannPotential1991}.

\section{Out of the Blue into the Middle of the Quantum Debate}

Looking again through my written records from that time at CERN I must say that I was totally fascinated and absorbed by the extraordinary personality of John Bell. I admired his knowledge and wisdom, for me he was \emph{the} man who had a deep understanding of quantum field theory and could simplify its issues within very concrete examples of potential theory. His maxim was ``\emph{Always test your general reasoning against simple models!}'' So it was quite understandable that I was working passionately in our collaboration.

Of course, I had heard that he was also a leading figure in quantum mechanics, specifically, in quantum foundations. But nobody could actually explain to me his work in this quantum area, neither at CERN nor anywhere else. The standard answer was: He discovered some `relation' whose consequence was that quantum mechanics turned out alright, but this we know anyhow, so don't worry. And I didn't. John, on the other hand, never mentioned his quantum works to me in the first years of our collaboration. Why? This I understood later on, John was reluctant to push somebody into a field that was quite unpopular at that time (see remarks in Sect.~\ref{sec:Bell experiments}).\\

At the end of the summer of 1980, I returned for some time to my home institute in Vienna to continue our collaboration on the gluon condensate potential from there. At that time, there was no internet, and it was a common practice to send preprints (typed manuscripts) of work prior to publication to all main physics institutions in the world. Also we in Vienna had such a preprint shelf where each week the new incoming preprints were exhibited.

\begin{figure}
\begin{center}
(a)\;\setlength{\fboxsep}{2pt}\setlength{\fboxrule}{0.8pt}\fbox{
\includegraphics[angle = 0, width = 50mm]{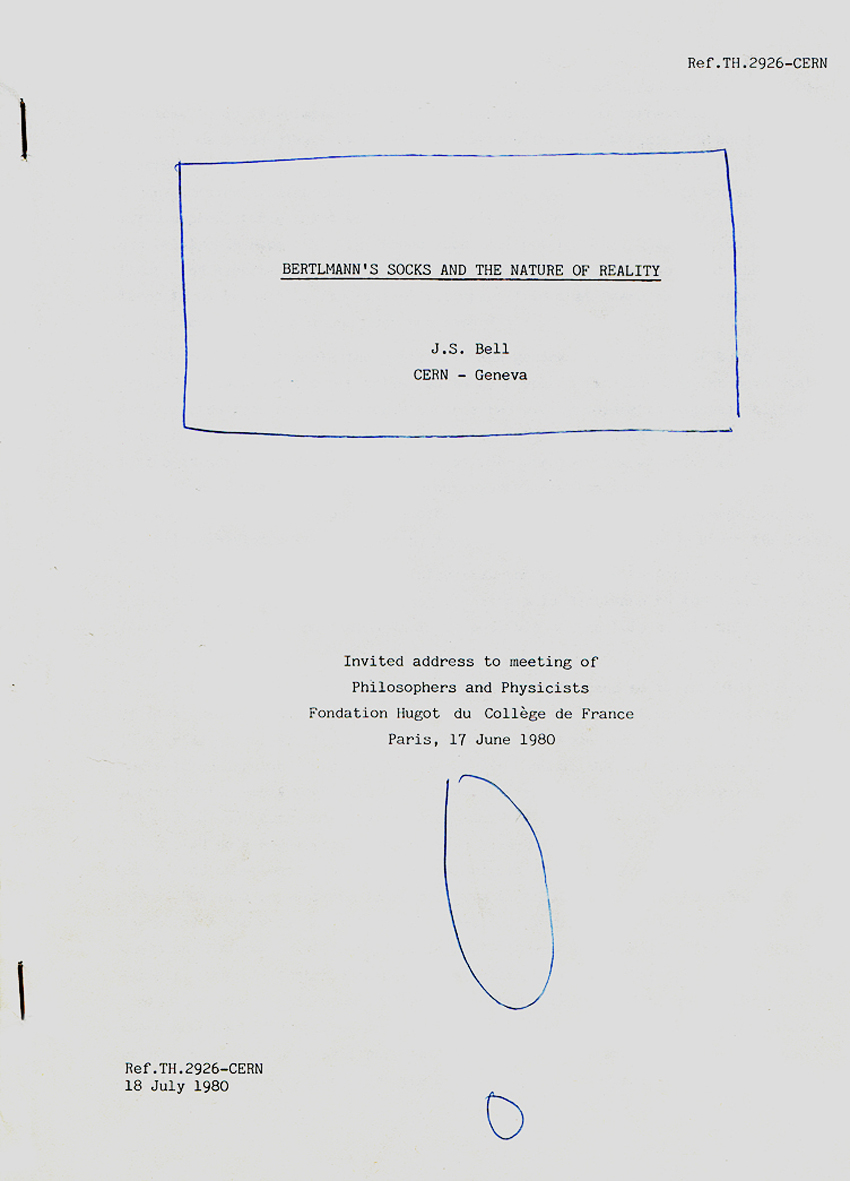}}
\hspace{5mm}
(b)\;\setlength{\fboxsep}{2pt}\setlength{\fboxrule}{0.8pt}\fbox{
\includegraphics[angle = 0, width = 50mm]{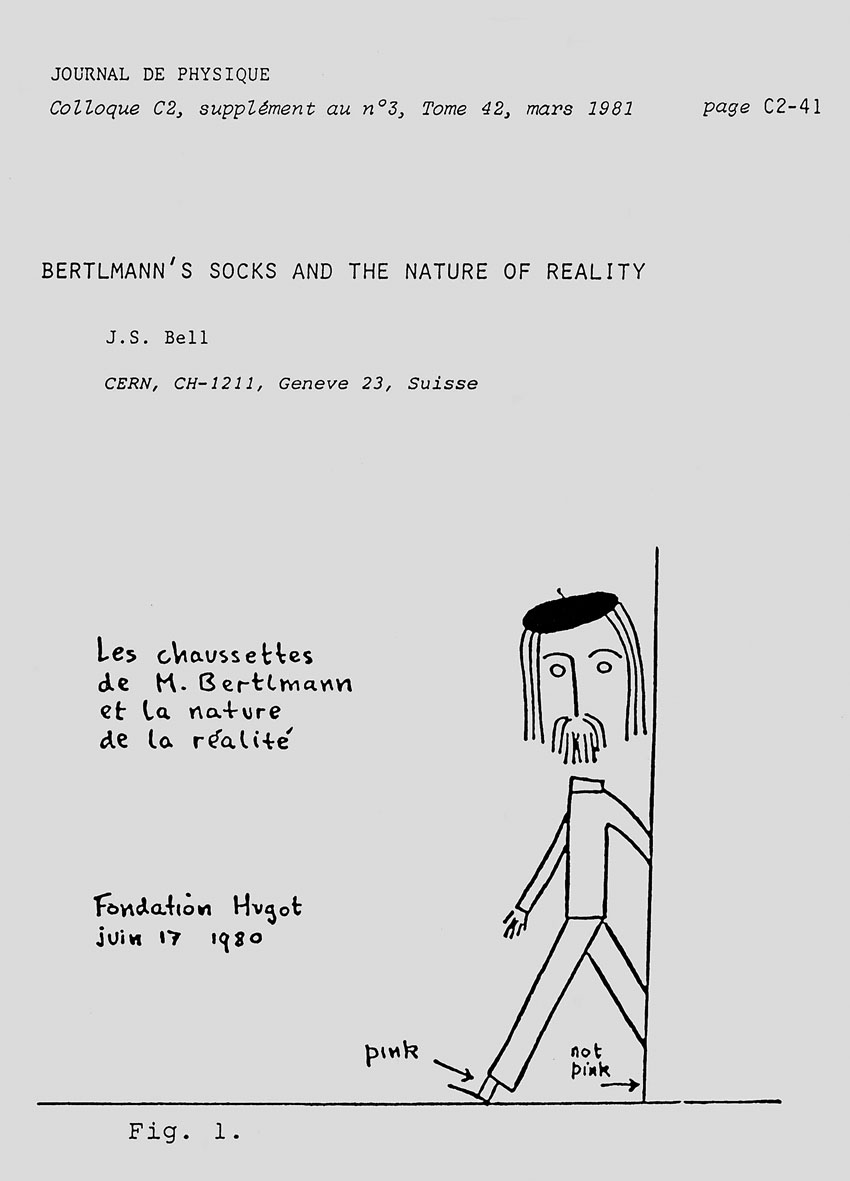}}
\normalsize{
\caption{(a) Original CERN preprint Ref.TH.2926-CERN \emph{``Bertlmann's socks and the nature of reality''} of John Bell from 18th July 1980 \cite{Bell-Bsocks-CERNpreprint}. It had been `decorated' by Gerhard Ecker who was in charge of receiving the preprints. (b) Cartoon about Bertlmann's socks published in Journal de Physique, Ref.~\cite{Bell-Bsocks-Journal-de-Physique}. The article is based on an invited lecture John Bell has given at le Colloque sur les \emph{``Implications conceptuelles de la physique quantique''}, organis\'{e} par la Foundation Hugot du Coll\`{e}ge de France, le 17 juin 1980.}
\label{fig:Bell-Bsocks-CERNpaper-cartoon}}
\end{center}
\end{figure}

One day, on the 15th of September, I was sitting in the institute's computer room, handling my computer cards, when my colleague Gerhard Ecker, who was in charge of receiving the preprints, rushed in waving a preprint in his hands. He shouted, ``\emph{Reinhold look---now you're famous!}'' I could hardly believe my eyes as I read and reread the title of a paper by John S. Bell~\cite{Bell-Bsocks-CERNpreprint} (see Fig.~\ref{fig:Bell-Bsocks-CERNpaper-cartoon}): ``\emph{Bertlmann's socks and the nature of reality}''.\\

I was totally excited. Reading the first page my heart stood still:\\

\emph{The philosopher in the street, who has not suffered a course in quantum mechanics, is quite unimpressed by Einstein-Podolsky-Rosen correlations \cite{EPR}. He can point to many examples of similar correlations in every day life. The case of Bertlmann's socks is often cited. Dr. Bertlmann likes to wear two socks of different colours. Which colour he will have on a given foot on a given day is quite unpredictable. But when you see (Fig.~\ref{fig:Bell-Bsocks-CERNpaper-cartoon}) that the first sock is pink you can be already sure that the second sock will not be pink. Observation of the first, and experience of Bertlmann, gives immediate information about the second. There is no accounting for tastes, but apart from that there is no mystery here. And is not the EPR business just the same ?}\\

Seeing the cartoon John has sketched himself showing me with my odd socks nearly knocked me down. All this came so unexpectedly---I had not the slightest idea that John had noticed my habits of wearing socks of different colours, a habit I cultivated since my early student days, my special generation-68 protest. This article pushed me \emph{instantaneously} into the quantum debate, which changed my life. Since then \emph{``Bertlmann's socks''} has developed a life of its own. You can find Bertlmann's socks everywhere on the internet, in popular science debates, and even in the fields of literature and art.\\

Now the time has come for diving into the quantum work of John, to understand his contributions in the quantum debate. It was John who pushed the rather philosophical Einstein-Bohr discussions of the 1930s about realism and incompleteness of quantum mechanics onto physical grounds. His axiom of locality or separability was the essential ingredient of a hidden variable theory and illuminated physical differences between all such hidden variable theories and the predictions of quantum mechanics. As result of \emph{``Bell's Theorem''} we can distinguish \emph{experimentally} between quantum mechanics and all local realistic theories with hidden variables. I was impressed by the clarity and depth of John's thoughts. From this time on we had fruitful discussions about the foundations of quantum mechanics and this was a great fortune and honour for me. It was just about the time when Alain Aspect \cite{Aspect-Dalibard-Roger1982} finished his time-flip experiments on Bell inequalities and the whole field began to attract the increasing interest of physicists.

For me a new world opened up---the universe of John Bell---and caught my interest and fascination for the rest of my life. In the following chapters I will present my personal view of this universe with the implications it had for my own research.

\section{Hidden Variables Theories -- Contextuality}

\subsection{Bell's Dissatisfaction with Quantum Mechanics}

Bell's dissatisfaction with quantum mechanics (QM) can be traced back to the time when he was a student at Queen's University in Belfast (1948 - 1949). In particular he disliked the so-called \emph{Copenhagen Interpretation} with its distinction between the quantum world and the classical world, the quantum system described by the wave function and the measuring apparatus as a classical device. He was thinking deeply about QM, not just how to use it, but about its conceptual meaning. He wondered \emph{``where does the quantum world stop and the classical world begin''}. So he always thought about how to get rid of that division.

For him it was clear that hidden variable theories would be appropriate to reformulate quantum theory, even though they were totally out of favour in the physics community. This shows Bell's strong moral character. If the quantum particles do \emph{have} definite properties, hidden variables, then we don't have to be concerned that the classical apparatus has definite properties. \emph{``Everything has definite properties!''} I remember John saying.

\subsection{Hidden Variable Theories}

Hidden variable theories (HVT) as well as quantum mechanics describe an ensemble of individual systems. Whereas in QM the orthodox (Copenhagen) doctrine tells us that measured properties, e.g. the spin of a particle, have no definite values before measurement, the HVT in contrast postulate that the properties of individual systems do \emph{have} pre-existing values revealed by the act of measurement.

What are the features of a HVT quite generally? Let us consider an ensemble of individual systems, each of which is prepared in a state $\ket{\psi}\,$, which is described by a set of observables
\begin{equation}\label{set-of-observables}
A, B, C, \ldots \quad\;.
\end{equation}
A HVT~\cite{Mermin-RevModPhys} assigns to each individual system a set of values corresponding to the observables (\ref{set-of-observables}), one of the eigenvalues of the corresponding operator
\begin{equation}\label{set-of-values-observables}
v(A), v(B), v(C), \ldots \quad\;,
\end{equation}
such that a measurement of observable $A$ in an individual system gives the numerical value $v(A)\,$.

The HVT now provides a rule how the values (\ref{set-of-values-observables}) should be distributed over all individual systems of the ensemble that is given by the state $\ket{\psi}\,$. Of course, it must be such that the statistical distribution of the results agrees with QM. The states, specified by the quantum mechanical state vector \emph{and} by an additional hidden variable which determines individually the results as in classical statistical mechanics, are called \emph{dispersion-free}.

If, in particular, a functional relation is satisfied
\begin{equation}\label{functional-relation-observables}
f(A, B, C, \ldots ) \;=\; 0 \;,
\end{equation}
by a set of mutually commuting observables $A, B, C, \ldots$ then the same relation must hold for the values in the individual systems
\begin{equation}\label{functional-relation-values-observables}
f\big(v(A), v(B), v(C), \ldots \big) \;=\; 0 \;.
\end{equation}

Amazingly, just by relying on conditions (\ref{functional-relation-observables}) and (\ref{functional-relation-values-observables}) one can construct so-called \emph{No-Go Theorems} that arrive at a contradiction. I will explain these theorems in the following sections.

\subsection{Von Neumann and Additivity of Measurement Values}

John Bell started his investigation \emph{``On the problem of hidden variables in quantum mechanics''} \cite{Bell-RevModPhys1966} by criticizing von Neumann. John von Neumann had written already in 1932 \cite{vonNeumann1932} a proof that dispersion-free states, and thus hidden variables, are incompatible with QM. Due to the high reputation von Neumann had in the physics community his proof was widly accepted. Bell, in contrast, examined von Neumann's proof carefully and critically. The essential point was the following.

Consider three operators with condition (\ref{functional-relation-observables})
\begin{equation}\label{operator-additivity}
C \;=\; A \,+\, B \;,
\end{equation}
then it follows that the corresponding attached values in the individual systems must also satisfy
\begin{equation}\label{value-operator-additivity}
v(C) \;=\; v(A) \,+\, v(B) \;,
\end{equation}
since the operators $A, B$ are supposed to commute.

Now von Neumann's assumption was to impose condition (\ref{value-operator-additivity}) on a hidden variable theory also for \emph{noncommuting} operators. Interestingly, already in 1935 the mathematician and philosopher Grete Hermann \cite{GreteHermann1935} raised her objection to von Neumann's assumption but she was totally ignored. Also Simon Kochen and Enst Specker \cite{Kochen2002}, when reading von Neumann's proof in 1961, had their doubts about the additivity (\ref{value-operator-additivity}) for noncommuting operators. And John Bell grumbled, \emph{``This is wrong!''}

Why? Consider the following example. The measurement of the $\sigma_x$ operator for a magnetic particle requires a suitably oriented Stern-Gerlach apparatus. The measurement of $\sigma_y$ demands a different orientation and $(\sigma_x + \sigma_y)$ again a different one. The operators cannot be measured simultaneously, thus there is no reason for imposing  the additivity relation (\ref{value-operator-additivity}). In fact, von Neumann was somehow misled by QM since for the quantum mechanical expectation values we have
\begin{equation}\label{operator-additivity-inthemean}
\langle \psi | A + B | \psi \rangle \;=\; \langle \psi | A | \psi \rangle \,+\, \langle \psi | B | \psi \rangle \;,
\end{equation}
that is, additivity holds in the mean, irrespective whether $A, B$ commute or not.

Then Bell proceeds to construct a simple 2-dimensional example demonstrating that von Neumann's additivity assumption (\ref{value-operator-additivity}) is not reasonable for noncommuting observables.

\subsection{Bell's 2-dimensional Hidden Variable Model}\label{Bell-HV-model}

Consider a spin measurement along some direction $\vec{n}$
\begin{equation}\label{spin-eigenvalue-equation}
\vec{\sigma}\cdot\vec{n} \,\ket{\pm\vec{n}} \;=\; \pm \ket{\pm\vec{n}} \;,
\end{equation}
where $\ket{\pm\vec{n}}$ are the eigenstates of operator $\vec{\sigma}\cdot\vec{n}$ measuring the spin. Each observable can be represented quite generally by Pauli matrices
\begin{equation}\label{observable-general-decompos}
A \;=\; a_0 \mathds{1} \,+\, \vec{a}\cdot\vec{\sigma} \;, \qquad B \;=\; b_0 \mathds{1} \,+\, \vec{b}\cdot\vec{\sigma} \;.
\end{equation}
The operators $A, B$ commute if and only if $\vec{a}$ and $\vec{b}$ are parallel or antiparallel. The allowed values for an individual system are
\begin{equation}\label{values-individually}
v(A) \;=\; a_0 \,\pm\, |\vec{a}| \;, \qquad v(B) \;=\; b_0 \,\pm\, |\vec{b}| \;.
\end{equation}
Therefore, if the observables commute---let us first choose the vectors $\vec{a}$ and $\vec{b}$ to be parallel---we obtain
\beq\label{values-additive}
v(A + B) &\;=\;& v\big( (a_0 + b_0) \mathds{1} \,+\, (\vec{a} + \vec{b})\cdot\vec{\sigma} \big) \nonumber\\
&\;=\;& a_0 + b_0 \,\pm\, |\vec{a} + \vec{b}| \nonumber\\
&\;=\;& v(A) + v(B) \;,
\eeq
the corresponding attached values are additive since $|\vec{a} + \vec{b}| \,=\, |\vec{a}| + |\vec{b}|\,$. If $\vec{a}$ and $\vec{b}$ are antiparallel we have $|\vec{a} + \vec{b}| \,=\, |\vec{a}| - |\vec{b}|$ for $|\vec{a}| > |\vec{b}|\, $, which leads to the same additivity result (\ref{values-additive}) since in this case $v(B) \;=\; b_0 \,\mp\, |\vec{b}|\,$.

If, however, the observables are noncommuting---the vectors are not (anti)parallel---then we have $|\vec{a} + \vec{b}| \,\neq\, |\vec{a}| + |\vec{b}|$ and consequently
\beq\label{values-nonadditive}
v(A + B) &\;\neq\;& v(A) + v(B) \;.
\eeq

Bell went on to show the invalidity of von Neumann's conclusion by formulating a hidden variable (HV) model that reproduces the QM predictions in this system. Consider the expectation value, for such a mean value we can distribute the single values in the following way
\beq\label{HV-distribution}
v(A) &\;=\;& a_0 \,+\, |\vec{a}| \qquad \mbox{for} \qquad (\vec{\lambda} + \vec{n})\cdot\vec{a} > 0     \nonumber\\
v(A) &\;=\;& a_0 \,-\, |\vec{a}| \qquad \mbox{for} \qquad (\vec{\lambda} + \vec{n})\cdot\vec{a} < 0 \;,
\eeq
where $\vec{\lambda}$ denotes some random unit vector, the hidden variable. Then the mean value of (\ref{HV-distribution}) over a uniform distribution of directions $\vec{\lambda}$ provides the quantum mechanical result as required
\begin{equation}\label{HV-mean-value}
\bar{v}(A) \;=\; \frac{1}{4\pi}\int\!d\Omega(\vec{\lambda})\,v(A) \;=\; a_0 \,+\, \vec{a}\cdot\vec{n} \;=\; \langle +\vec{n}|\,A\,|+\vec{n}\rangle \;,
\end{equation}
and, of course, with the additivity property
\begin{equation}\label{additivity-mean-value}
\langle +\vec{n}|\,A + B\,|+\vec{n}\rangle \;=\; \langle +\vec{n}|\,A\,|+\vec{n}\rangle \,+\, \langle +\vec{n}|\,B\,|+\vec{n}\rangle \;.
\end{equation}
This is a simple 2-dimensional HV model for QM demonstrating that von Neumann's assumption (\ref{value-operator-additivity}) is not justified.

\begin{figure}
\centering
\includegraphics[width=0.3\textwidth]{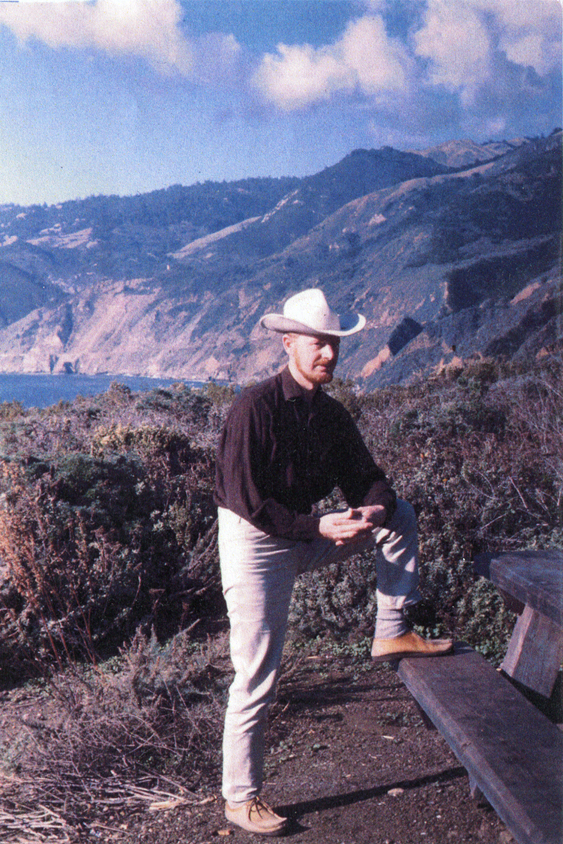}
\caption{John Bell presenting himself as real Californian in 1964, while spending his sabbatical at SLAC, USA. It was the time when he worked on hidden variables and wrote his celebrated paper on Bell's inequality. Foto: \copyright Mary Bell.}
	\label{fig:Bell-USA-1964}
\end{figure}

\subsection{Gleason and Contextuality}

Josef-Maria Jauch, professor for theoretical physics at the University of Geneva, drew Bell's attention to the work of Andrew M. Gleason. The rather mathematical work of Gleason \cite{Gleason1957} was not explicitly addressed to HVT but aimed instead to reduce the axioms for QM. The theorem of Gleason states the following:

\begin{theorem}[Gleason's Theorem]\ \
In a Hilbert space $\Ha$ of $dim \,\Ha > 2$ the only probability measures of the state associated with a linear subspace $\V$ of the Hilbert space are of the form $\;\tr \,P(\V)\rho\,$, where $P(\V)$ is the projection operator onto $\V$ and $\rho$ the density matrix of the system.
\label{theorem:Gleason}
\end{theorem}

A property of Theorem~\ref{theorem:Gleason} is certainly that the probability of the sum of commuting projections equals the sum of the probabilities of those projections individually. However, $\;\tr \,P(\V)\rho\,$ cannot have values restricted just to $0$ or $1$ for all projection operators $P(\V)\,$, and consequently the state represented by $\rho$ cannot be dispersion-free for each observable.

Always preferring to construct his own proof, Bell established the following corollary~\cite{Bell-RevModPhys1966}. It is more directed to HVT.

\begin{corollary}[Bell's Corollary]\ \
Consider a state space $\V$. If $dim \,\V > 2$ then the additivity requirement for expectation values of commuting operators cannot be met for dispersion-free states.
\label{corollary:Bell}
\end{corollary}

Corollary~\ref{corollary:Bell} states that for $dim \V > 2$ it is in general impossible to assign a definite value for each observable in each individual quantum system. Note that this is not in conflict with Bell's HV model (Sect.~\ref{Bell-HV-model}), which has only two dimensions.\\

In his HV paper, which Bell wrote in 1964, while spending his sabbatical at SLAC, USA (see Fig.~\ref{fig:Bell-USA-1964}), Bell formulated the impossibility to assign a definite value for each observable in each individual system in the following way~\cite{Bell-RevModPhys1966}:\\

\emph{``It was tacitly assumed that measurement of an observable must yield the same value independently of what other measurements may be made simultaneously. Thus as well as $P(\phi_3)$ say, one might measure either $P(\phi_2)$ or $P(\psi_2)$, where $\phi_2$ and $\psi_2$ are orthogonal to $\phi_3$ but not to one another. These different possibilities require different experimental arrangements; there is no a priori reason to believe that the results for $P(\phi_3)$ should be the same. The result of an observation may reasonably depend not only on the state of the system (including hidden variables) but also on the complete disposition of the apparatus.''}\\

Thus Bell pointed to another class of hidden variable models, where the results may depend on different settings of the apparatus. Such models are called \emph{contextual} and may agree with quantum mechanics. Corollary~\ref{corollary:Bell}, on the other hand, states that all \emph{noncontextual} HVT are in conflict with QM (for $dim > 2$). Hence the essential feature for the difference between HVT and QM is \emph{contextuality}.

In 1967, Simon Kochen and Ernst Specker published their famous paper on \emph{``The problem of hidden variables in quantum mechanics''} \cite{Kochen-Specker}, where they established their no-go theorem that noncontextual hidden variable theories are incompatible with quantum mechanics.

Since then, contextuality has become an important issue in the research of quantum systems (see, e.g., Refs.~\cite{Cabello-Guehne-etal_PRL2013, Canas-Cabello-etal2013, Cabello-etal_PRL-112-040401-2014, Cabello-etal_PRL-111-180404-2013, Guehne-Cabello-etal2013, Acin-etal2012}, and references therein).

\subsection{Bohm and Nonlocality}

In 1952, David Bohm~\cite{Bohm1952} published his vision of how to reinterpret quantum theory as a deterministic, realistic theory with hidden variables. His work was not appreciated by the physics community, not even by Einstein, with whom Bohm had discussed these issues. One might think that Einstein would have been particularly enthusiastic about Bohm's view. Not at all. In a letter~\cite{Einstein-Born-letter} to his friend Max Born dated May 12th, 1952, Einstein dismissed Bohm's theory, saying, \emph{``That way seems too cheap to me.''} This was partly because Bohm had to introduce additional parameters in order to save realism and determinism and partly because his model did not contain any new physics, it was just equivalent to quantum mechanics. Also Wolfgang Pauli rejected Bohm's work as \emph{``artificial metaphysics''}. John Bell, however, was very much impressed by Bohm's work and often remarked, \emph{``I saw the impossible thing done''.} To me John continued, \emph{``In every quantum mechanics course you should learn Bohm's model!''}\\

At the end of his HV paper~\cite{Bell-RevModPhys1966}, Bell examined Bohm's model quite carefully. He considered a system of two particles with spin $\frac{1}{2}\,$. The quantum mechanical state is described by a wave function
\begin{equation}\label{wave-function}
\psi_{ij}(\vec{r}_1, \vec{r}_2) \;,
\end{equation}
with $i, j$ for the spin indices. The wave function is governed by the Schr\"odinger equation
\beq\label{Schroedinger-equation}
i\hbar\,\frac{d}{dt} \,\psi_{ij}(\vec{r}_1, \vec{r}_2) &\;=\;& \Big[ -\frac{\hbar^2}{2m} \big( \frac{\partial^2}{\partial \vec{r}_1^{\,2}} \,+\, \frac{\partial^2}{\partial \vec{r}_2^{\,2}} \big) \,+\, V(\vec{r}_1 - \vec{r}_2) \nonumber \\
&\;+\;& \mu_1 \,\vec{\sigma}_1\cdot\vec{B}(\vec{r}_1) \,+\,  \mu_2 \,\vec{\sigma}_2\cdot\vec{B}(\vec{r}_2) \Big] \,\psi_{ij}(\vec{r}_1, \vec{r}_2) \;,
\eeq
where $V$ represents the interaction potential and $\vec{B}$ the external magnetic field of the magnets that analyze the spins.

The hidden variables are two vectors $\vec{X}_1$ and $\vec{X}_2$ which yield the results for position measurements. The variables are supposed to be distributed in configuration space with probability density
\begin{equation}\label{probability-density}
\rho (\vec{X}_1, \vec{X}_2) \;=\; \sum_{i,j} |\psi_{ij}(\vec{X}_1, \vec{X}_2)|^2 \;,
\end{equation}
which describes the quantum mechanical state.

For a one-particle system the position operator would follow the time evolution
\begin{equation}\label{position-time-evolution-general}
\frac{d\vec{X}}{dt} \;=\; \frac{\vec{j}(\vec{X},t)}{\rho (\vec{X},t)} \;,
\end{equation}
where $\vec{j}$ denotes the probability current calculated in the usual way
\begin{equation}\label{probability-current}
\vec{j}(\vec{X},t) \;=\; \frac{\hbar}{2m} \,{\rm{Im}}\, \psi^\ast (\vec{X},t)\frac{\partial}{\partial\vec{X}}\psi (\vec{X},t) \;.
\end{equation}
For our two-particle system the position operators, the hidden variables, then vary in time according to ($\hbar = 1, 2m = 1$)
\beq\label{position-operators-time-evolution}
\frac{d\vec{X}_1}{dt} &\;=\;& \frac{1}{\rho (\vec{X}_1,\vec{X}_2,t)} \,{\rm{Im}}\, \sum_{i,j} \psi_{ij}^\ast (\vec{X}_1,\vec{X}_2,t)\frac{\partial}{\partial\vec{X}_1}\psi_{ij} (\vec{X}_1,\vec{X}_2,t) \nonumber\\
\frac{d\vec{X}_2}{dt} &\;=\;& \frac{1}{\rho (\vec{X}_1,\vec{X}_2,t)} \,{\rm{Im}}\, \sum_{i,j} \psi_{ij}^\ast (\vec{X}_1,\vec{X}_2,t)\frac{\partial}{\partial\vec{X}_2}\psi_{ij} (\vec{X}_1,\vec{X}_2,t) \;.
\eeq
The strange feature now is that the trajectory equations (\ref{position-operators-time-evolution}) for the operators, the hidden variables, have a highly nonlocal character. Only in case of a factorizable wave function for the quantum system
\begin{equation}\label{wave-function-factorizable}
\psi_{ij}(\vec{X}_1, \vec{X}_2, t) \;=\; \eta_i (\vec{X}_1,t) \cdot \chi_j (\vec{X}_2,t)  \;,
\end{equation}
the trajectories decouple
\beq\label{position-operators-time-evolution-decoupled}
\frac{d\vec{X}_1}{dt} &\;=\;& \frac{1}{\sum_{i} |\eta_i (\vec{X}_1,t)|^2} \,{\rm{Im}}\, \sum_{i} \eta_{i}^\ast (\vec{X}_1,t)\frac{\partial}{\partial\vec{X}_1}\eta_i (\vec{X}_1,t) \nonumber\\
\frac{d\vec{X}_2}{dt} &\;=\;& \frac{1}{\sum_{j} |\chi_j (\vec{X}_2,t)|^2} \,{\rm{Im}}\, \sum_{j} \chi_{j}^\ast (\vec{X}_2,t)\frac{\partial}{\partial\vec{X}_2}\chi_j (\vec{X}_2,t) \;.
\eeq
The Schr\"odinger equation (\ref{Schroedinger-equation}) separates too and the trajectories of $\vec{X}_1$ and $\vec{X}_2$ are determined separately by involving the magnetic fields $\vec{B}(\vec{X}_1)$ and $\vec{B}(\vec{X}_2)$ respectively. However, in general this is not the case. The particle $1$ depends in a complicated way on the trajectory and wave function of particle $2\,$, no matter how remote the particles are.\\

Bohm was aware of the peculiar feature of his model that particle $1$ depends on the characteristics of particle $2\,$, but it was Bell who realized the importance of it. Bell wondered if it was just a defect of this particular HV model, or is it somehow intrinsic in a hidden variable theory reproducing quantum mechanics. Bell was also acquainted with Bohm's spin version~\cite{Bohm-Aharanov-EPRspin1957} of the Einstein-Podolsky-Rosen (EPR) paradox, where distant correlations were considered. This two-spin setup \`a la EPR he investigated further.\\

John told me once:

\emph{``At the beginning I just played around to get simple relations which would give a local account for the quantum correlations but everything I tried didn't work. So I felt it couldn't be done and then I constructed an impossibility proof.''}\\

Historically, Bell's impossibility proof for local hidden variables~\cite{Bell-Physics1964}, local realistic theories, was published already in 1964, before Bell could present his earlier work on hidden variables~\cite{Bell-RevModPhys1966}. The reason was that the HV work remained unattended in a drawer of the editorial office of Review of Modern Physics. SLAC's post office was not efficient enough to forward the editorial reply to Bell, who moved meanwhile to an other university in the USA. Only when Bell, getting no response for a year, inquired politely about the status of his paper, the editorial office reconsidered it and published it in 1966.

\section{Bell Inequalities -- Nonlocality}

\subsection{Bell's Locality Hypothesis}

Bell's starting point was: \emph{``On the Einstein-Podolsky-Rosen paradox''}~\cite{Bell-Physics1964}. More precisely, he considered Bohm's spin version~\cite{Bohm-Aharanov-EPRspin1957} of it. The paradox of EPR served as an argument that quantum mechanics is an incomplete theory and that it should be supplemented by additional parameters, the hidden variables. These additional variables would restore causality and locality in the theory.

Bell's profound discovery was that the requirement of locality is \emph{incompatible} with the statistical predictions of quantum mechanics. He phrased the requirement that created the essential difficulty in the following way~\cite{Bell-Physics1964}:\\

\emph{``The result of a measurement on one system be unaffected by operations on a distant system with which it has interacted in the past.''}\\

\begin{figure}
\centering
\includegraphics[width=0.65\textwidth]{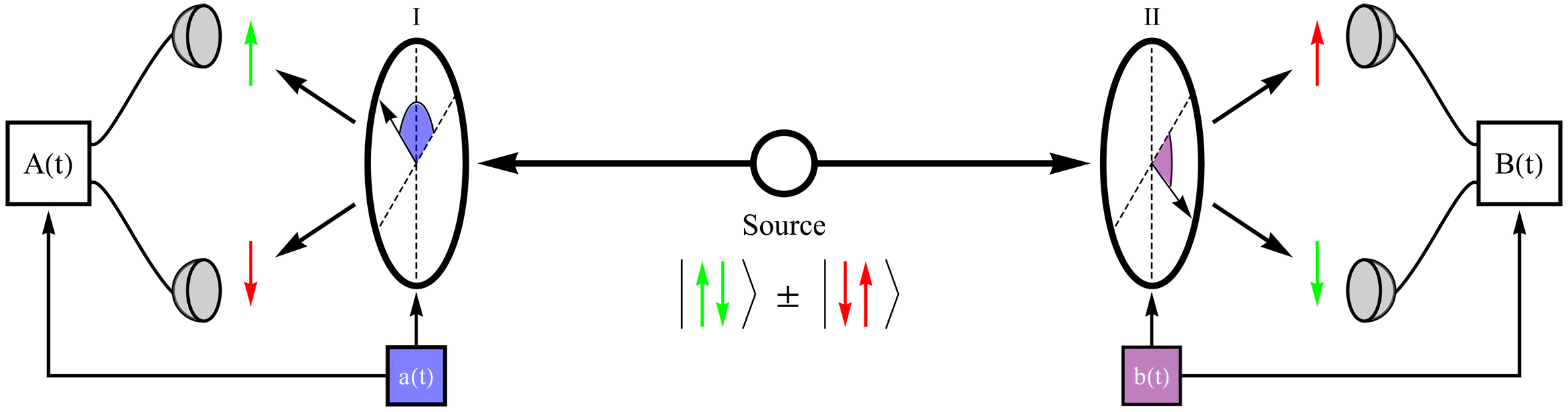}
\caption{In a Bohm-EPR setup a pair of spin $\frac{1}{2}$ particles, prepared in a spin singlet state, propagates freely in opposite directions to the measuring stations called Alice and Bob. Alice measures the spin in direction $\vec{a}$, whereas Bob measures simultaneously in direction $\vec{b}$.}
	\label{fig:Bell-setup-experiment}
\end{figure}

In such a Bohm-EPR setup a pair of spin $\frac{1}{2}$ particles is produced in a spin singlet state and propagates freely into opposite directions (see Fig.~\ref{fig:Bell-setup-experiment}).
The spin measurement on one side, called Alice, performed by a Stern-Gerlach magnet along some direction $\vec{a}$ is described by the operator $\vec{\sigma}_A \cdot \vec{a}\,$ and yields the values $\pm 1\,$. Since we can predict in advance the result of $\vec{\sigma}_B \cdot \vec{b}\,$ on the other side, Bob's side, the result must be predetermined. This predetermination we specify by the additional variable $\lambda\,$. In such an extended theory we denote the measurement result of Alice and Bob by
\begin{equation}
A(\vec{a},\lambda) \;=\; \pm 1,0 \qquad \mbox{and} \qquad B(\vec{b},\lambda) \;=\; \pm 1,0 \;.
\end{equation}
We also include $0$ for imperfect measurements to be more general, i.e., what we actually require is
\begin{equation}
|A| \;\leq\; 1 \qquad \mbox{and} \qquad |B| \;\leq\; 1 \;.
\end{equation}
Then the expectation value of the joint spin measurement of Alice and Bob is
\begin{equation}\label{expectation-value-Bell-locality}
E(\vec{a},\vec{b}) \;=\; \int\!d\lambda\,\rho(\lambda)\,A(\vec{a},\lambda)\cdot B(\vec{b},\lambda) \;.
\end{equation}
This choice of the product $A(\vec{a},\lambda)\cdot B(\vec{b},\lambda)$ in expectation value (\ref{expectation-value-Bell-locality}) is called \emph{Bell's locality hypothesis}. It is the obvious definition of a physicist as an engineer, and must not be confused with locality definitions in quantum field theory.

The function $\rho(\lambda)$ represents some probability distribution for the variable $\lambda$, and does not depend on the measurement settings $\vec{a}$ and $\vec{b}\,$, which can be chosen truly free or random. This is essential! The distribution is normalized
\begin{equation}\label{normalization-of-rho}
\int\!d\lambda\,\rho(\lambda) \;=\; 1 \;.
\end{equation}

Now it is quite easy to derive an inequality that is well adapted to real experiments with apparatus inefficiencies. Starting with the following difference of expectation values
\beq\label{difference-expectation-values}
E(\vec{a},\vec{b}) \,-\, E(\vec{a},\vec{b}^{\,'}) &\;=\;& \int\!d\lambda\,\rho(\lambda)\,A(\vec{a},\lambda)\cdot B(\vec{b},\lambda)
\big( 1 \,\pm\, A(\vec{a}^{\,'},\lambda)\cdot B(\vec{b}^{\,'},\lambda) \big) \nonumber\\
&\;-\;& \int\!d\lambda\,\rho(\lambda)\,A(\vec{a},\lambda)\cdot B(\vec{b}^{\,'},\lambda)
\big( 1 \,\pm\, A(\vec{a}^{\,'},\lambda)\cdot B(\vec{b},\lambda) \big) \;,
\eeq
taking the absolute values
\beq\label{absolute-difference-expectation-values}
|E(\vec{a},\vec{b}) \,-\, E(\vec{a},\vec{b}^{\,'})| &\;\le\;& \int\!d\lambda\,\rho(\lambda)
\big( 1 \,\pm\, A(\vec{a}^{\,'},\lambda)\cdot B(\vec{b}^{\,'},\lambda) \big) \nonumber\\
&\;+\;& \int\!d\lambda\,\rho(\lambda)
\big( 1 \,\pm\, A(\vec{a}^{\,'},\lambda)\cdot B(\vec{b},\lambda) \big) \nonumber\\
&\;\le\;& 2 \,\pm\, |E(\vec{a}^{\,'},\vec{b}^{\,'}) \,+\, E(\vec{a}^{\,'},\vec{b})| \;,
\eeq
and choosing the minus sign in Eq.~(\ref{absolute-difference-expectation-values}), which makes the inequality tighter, we find
\beq\label{CHSH-inequality}
S_{\rm{CHSH}} \;:=\; |E(\vec{a},\vec{b}) \,-\, E(\vec{a},\vec{b}^{\,'})| \,+\, |E(\vec{a}^{\,'},\vec{b}) \,+\, E(\vec{a}^{\,'},\vec{b}^{\,'})| \;\le\; 2 \;.
\eeq
Inequality (\ref{CHSH-inequality}) is the familiar \emph{CHSH Inequality}, named after Clauser, Horne, Shimony, and Hold who published it in 1969 \cite{CHSH}. Bell presented his own, more general derivation of (\ref{CHSH-inequality}) at the International School of Physics ``Enrico Fermi'' in Varenna at Lake Como in 1970 \cite{Bell-Fermi-School-1971} and in his \emph{``Bertlmann's socks''} paper in 1980 \cite{Bell-Bsocks-CERNpreprint, Bell-Bsocks-Journal-de-Physique}.

Calculating now the quantum mechanical expectation value for the joint measurement when the system is in the spin singlet state $\ket{\psi^{\,-}} \,=\, \frac{1}{\sqrt{2}}(\ket{\Uparrow} \otimes \ket{\Downarrow} \,-\, \ket{\Uparrow} \otimes \ket{\Downarrow})\,$, also called Bell state,
\beq\label{expectation-value-QM-joint-measurement}
E(\vec{a},\vec{b}) &\;=\;& \langle \psi^{\,-} | \,\vec{a} \cdot \vec{\sigma}_A \otimes \vec{b} \cdot \vec{\sigma}_B \,| \psi^{\,-} \rangle \nonumber\\
&\;=\;& -\, \vec{a}\cdot \vec{b} \;=\; -\, \cos(\alpha - \beta)\;,
\eeq
where $\alpha,\beta$ are the angles of the orientations in Alice's and Bob's parallel planes, then we know that for the choice of the \emph{``Bell angles''} $(\alpha , \beta , \alpha^{\,'} , \beta^{\,'}) \;=\; (0, \frac{\pi}{4}, 2 \frac{\pi}{4}, 3 \frac{\pi}{4})$ the CHSH inequality (\ref{CHSH-inequality}) is maximally violated
\beq\label{CHSH-max-violation}
S_{\rm{CHSH}}^{\rm{QM}} \;=\; 2 \sqrt{2} \;=\; 2.828 \;>\; 2 \,.
\eeq
Inequality (\ref{CHSH-inequality}) had been tested experimentally by Zeilinger's group by using entangled photons in the Bell state $\ket{\psi^{\,-}}\,$ (see Sect.~\ref{third-generation-experiments}). In the photon case, the expectation value (\ref{expectation-value-QM-joint-measurement}) changes to $E(\vec{a},\vec{b}) \,=\, -\, \cos2(\alpha - \beta)\,$, i.e., the Bell angles become a factor of $2$ smaller as compared to the spin case.\\

In order to arrive next at Bell's original inequality we assume perfect (anti-)correlation
\begin{equation}
E(\vec{a},\vec{a}) \;=\; -1 \;,
\end{equation}
and choose only $3$ different orientations, which is the minimal choice, e.g., we equate $\vec{a}^{\,'} = \vec{b}^{\,'}\,$, then inequality (\ref{CHSH-inequality}) gives
\beq\label{Bells-inequality}
S_{\rm{Bell}} \;:=\; |E(\vec{a},\vec{b}) \,-\, E(\vec{a},\vec{b}^{\,'})| \,-\, E(\vec{b}^{\,'},\vec{b}) \;\le\; 1 \;,
\eeq
\emph{Bell's original inequality}~\cite{Bell-Physics1964}. Inequality (\ref{Bells-inequality}) is violated maximally for the quantum mechanical expectation value (\ref{expectation-value-QM-joint-measurement}) of the Bell state $\ket{\psi^{\,-}}$ and the choice of $(\alpha , \beta , \beta^{\,'}) \,=\, (0, 2\frac{\pi}{3}, \frac{\pi}{3})$ for the Bell angles
\beq\label{Bells-inequality-max-violation}
S_{\rm{Bell}}^{\rm{QM}} \;=\; \frac{3}{2} \;=\; 1.5 \;>\; 1 \,.
\eeq

\vspace{3mm}

When I studied and derived Bell's inequality (\ref{Bells-inequality}) for the first time, I was surprised and fascinated that quantum mechanics contradicted an inequality that relied on such general and quite `natural' assumptions. Bell's talent to turn the pure philosophical debate of Einstein and Bohr into exact mathematical terms was impressive. This formulation could be tested experimentally! It was amazing, how John was able to find this special linear combination of expectation values, which were in contradiction to QM at certain angles, called by himself the \emph{``awkward Irish angles''}. I got the feeling that there is something deep in it. Especially, because at that time, early 1980s, the experiments of Aspect received much attention, in particular his time-flip experiment \cite{Aspect-Dalibard-Roger1982} that revealed the nonlocal structure of Nature. But I certainly could not anticipate what was following afterwards, this explosion into a new area of quantum physics called nowadays quantum information, quantum communication and quantum computation \cite{nielsen-chuang2000, bertlmann-zeilinger02}.\\

Finally, I would like to discuss two other types of Bell inequalities which are often used in experiments. The first one has a very simple form and has been derived by Eugene P. Wigner in 1970~\cite{Wigner1970}. He focused on probabilities which are proportional to the number of clicks in a detector. In terms of probabilities $P$ for the joint measurements the expectation value can be expressed by
\beq\label{expectation-value-intermsof-probabilities}
E(\vec{a},\vec{b}) \;=\; P(\vec{a}\Uparrow,\vec{b}\Uparrow) \,+\, P(\vec{a}\Downarrow,\vec{b}\Downarrow) \,-\, P(\vec{a}\Uparrow,\vec{b}\Downarrow) \,-\, P(\vec{a}\Downarrow,\vec{b}\Uparrow)\,,
\eeq
and assuming that $P(\vec{a}\Uparrow,\vec{b}\Uparrow) \equiv P(\vec{a}\Downarrow,\vec{b}\Downarrow)$ and $P(\vec{a}\Uparrow,\vec{b}\Downarrow) \equiv P(\vec{a}\Downarrow,\vec{b}\Uparrow)$ together with $\sum P = 1$ the expectation value becomes
\beq\label{expectation-value-probability}
E(\vec{a},\vec{b}) \;=\; 4 \,P(\vec{a}\Uparrow,\vec{b}\Uparrow) \,-\, 1 \;.
\eeq
Inserting expression (\ref{expectation-value-probability}) into Bell's inequality (\ref{Bells-inequality}) yields \emph{Wigner's inequality} for the joint probabilities, where Alice measures spin $\Uparrow$ in direction $\vec{a}$ and Bob also $\Uparrow$ in direction $\vec{b}$ (we drop from now on the spin notation $\Uparrow$ in the formulae)
\beq\label{Wigners-inequality}
P(\vec{a},\vec{b}) \;\leq\; P(\vec{a},\vec{b}^{\,'}) \,+\, P(\vec{b}^{\,'},\vec{b}) \;,
\eeq
or rewritten
\beq\label{Wigners-inequality-for-S}
S_{\rm{Wigner}} \;:=\; P(\vec{a},\vec{b}) \;-\; P(\vec{a},\vec{b}^{\,'}) \,-\, P(\vec{b}^{\,'},\vec{b}) \;\leq\; 0 \;.
\eeq

For the Bell state $\ket{\psi^{\,-}}$ the quantum mechanical probability gives
\beq\label{joint-probability-for-Bell-state-psi-minus}
P(\vec{a},\vec{b}) \;=\; | \big(\bra{\vec{a}\Uparrow} \otimes \bra{\vec{b}\Uparrow} \big) \ket{\psi^{\,-}} |^2 \;=\; \frac{1}{2}\,\sin^2 \frac{1}{2}(\alpha - \beta) \;,
\eeq
and leads to a maximal violation of inequality (\ref{Wigners-inequality-for-S})
\beq\label{Wigner-inequality-max-violation}
S_{\rm{Wigner}}^{\rm{QM}} \;=\; \frac{1}{8} \;=\; 0.125 \;>\; 0 \,,
\eeq
for $(\alpha , \beta , \beta^{\,'}) \,=\, (0, 2\frac{\pi}{3}, \frac{\pi}{3})\,$, the same choice as in Bell's original inequality.\\

The last inequality I want to mention is the Clauser-Horne inequality of 1974 \cite{Clauser-Horne1974}. It relies on weaker assumptions and is very well suited for photon experiments with absorptive analyzers. Clauser and Horne work with relative counting rates, i.e., number of registered particles in the detectors. More precisely, the quantity $ N(\vec{a},\vec{b})$ is the rate of simultaneous events, coincidence rate, in the photon detectors of Alice and Bob after the photons passed the corresponding polarizers in direction $\vec{a}$ or $\vec{b}$ respectively. The relative rate $N(\vec{a},\vec{b})/N \,=\, P(\vec{a},\vec{b})\,$, where $N$ represents all events when the polarizers are removed, corresponds in the limit of infinitely many events, which is practically the case, to the joint probability $P(\vec{a},\vec{b})\,$. If one polarizer is removed, say on Bob's side, then expression $N_{A}(\vec{a})/N \,=\, P_{A}(\vec{a})$ stands for the single probability at Alice's (or the correspondingly at Bob's) side.

Starting from a pure algebraic inequality $-XY \;\leq\; x_1y_1 - x_1y_2 + x_2y_1 +x_2y_2 - Yx_2 - Xy_1 \,=\, S \;\leq\; 0$ for numbers $0 \,\leq\, x_1,x_2 \,\leq\, X$ and $0 \,\leq\, y_1,y_2 \,\leq\, Y\,$, it is now easy to derive the corresponding inequality for probabilities, which is the \emph{Clauser-Horne inequality}
\beq\label{Clauser-Horne-inequality}
-1 \;=\; P(\vec{a},\vec{b}) \,-\, P(\vec{a},\vec{b}^{\,'}) \,+\, P(\vec{a}^{\,'},\vec{b}) \,+\, P(\vec{a}^{\,'},\vec{b}^{\,'})
\,-\, P_{A}(\vec{a}^{\,'}) \,-\, P_{B}(\vec{b}) \;:=\; S_{\rm{CH}} \;\leq\; 0 \;.
\eeq

Inequality (\ref{Clauser-Horne-inequality}) has been used by Aspect in his time-flip experiment \cite{Aspect-Dalibard-Roger1982} (see Sect.~\ref{second-generation-experiments}). The two-photon state produced was the symmetrical Bell state $\ket{\phi^{\,+}} \,=\, \frac{1}{\sqrt{2}}(\ket{R}\otimes\ket{L} + \ket{L}\otimes\ket{R}) \,=\, \frac{1}{\sqrt{2}}(\ket{H}\otimes\ket{H} + \ket{V}\otimes\ket{V})\,$, where $\ket{R}, \ket{L}$ denote the right and left handed circularly polarized photons and $\ket{H}, \ket{V}$ the horizontally and vertically polarized ones.

In case of $\ket{\phi^{\,+}}$ entangled photons the quantum mechanical probability to detect a linear polarized photon with an angle $\alpha$ on Alice's side, and simultaneously an other linear polarized one with angle $\beta$ on Bob's side, is given by
\beq\label{joint-probability-for-Bell-state-phi-plus}
P(\vec{a},\vec{b}) \;=\; \big| \big[ \big( \bra{H}\cos\alpha \,+\, \bra{V}\sin\alpha \big) \otimes \big(\bra{H}\cos\beta \,+\, \bra{V}\sin\beta \big) \big] \ket{\phi^{\,+}} \big|^2 \;=\; \frac{1}{2}\,\cos^2 (\alpha - \beta) \;.
\eeq
Choosing now for the Bell angles $(\alpha , \beta , \alpha^{\,'} , \beta^{\,'}) \;=\; (0, \frac{\pi}{8}, 2 \frac{\pi}{8}, 3 \frac{\pi}{8})$ the quantum mechanical probabilities (\ref{joint-probability-for-Bell-state-phi-plus}) violate the Clauser-Horne inequality (\ref{Clauser-Horne-inequality}) maximally
\beq\label{Clauser-Horne-inequality-max-violation}
S_{\rm{CH}}^{\rm{QM}} \;=\; \frac{\sqrt{2} - 1}{2} \;=\; 0.207 \;>\; 0 \,.
\eeq

For further literature I want to refer to the review article \cite{Brunner-etal-Bell-nonlocality2013}.

\subsection{Conclusions}

What are the conclusions? In all Bell inequalities the essential ingredient is Bell's locality hypothesis, Eq.~(\ref{expectation-value-Bell-locality}), i.e., Einstein's vision of reality and Bell's concept of locality, therefore we have to conclude:

\begin{center}
\emph{Local realistic theories are incompatible with quantum mechanics!}
\end{center}

Bell in his seminal work~\cite{Bell-Physics1964}  realized the far reaching consequences of a realistic theory as an extension to quantum mechanics and expressed it in the following way:\\

\emph{``In a theory in which parameters are added to quantum mechanics to determine the results of individual measurements, without changing the statistical predictions, there must be a mechanism whereby the setting of one measuring device can influence the reading of another instrument, however remote. Moreover, the signal involved must propagate instantaneously, so that such a theory could not be Lorentz invariant.''}\\

He continued and stressed the crucial point in such EPR-type experiments:
\emph{``Experiments ... , in which the settings are changed during the flight of the particles, are crucial.''}

Thus it is of utmost importance \emph{not} to allow some mutual report by the exchange of signals with velocity less than or equal to that of light.

\section{Entanglement}

\subsection{Werner States}

In the 1980s, there was still the common belief that the violation of a Bell inequality, the nonlocal feature of QM, would imply that the quantum states involved were entangled, and vice versa. It was a great surprise, when in 1989 Reinhard Werner \cite{werner89} discovered that a certain mixture of entangled states may also satisfy a Bell inequality, i.e., behaves strictly local. Thus entanglement and nonlocality are not the same but \emph{different} concepts! How come?

Let us consider a bipartite quantum system given by its density matrix $\rho$ in the Hilbert-Schmidt space $\widetilde{\cal H} \,=\, {\widetilde{\cal H}}_A \otimes \widetilde{{\cal H}}_B$ of linear operators (also denoted by $L(\cal H)$) on the finite dimensional bipartite Hilbert space $\cal H \,=\, {\cal H}_A \otimes {\cal H}_B$ of Alice and Bob, with dimension $D = d_A \times d_B$. For our discussion of qubits $d_A = d_B = 2\,$. The quantum states $\rho$, the density matrices, are elements of $\widetilde{\cal H}$ with the properties $\rho^\dag = \rho$, Tr $\rho = 1$ and $\rho \geq 0$. A scalar product on $\widetilde{\cal H}$ is defined by $\left\langle A|B \right\rangle = \textnormal{Tr}\, A^\dag B$ with $A,B \in \widetilde{\cal H}$ and the corresponding squared norm is $\|A\|^2=\rm{Tr}\,A^\dag A$.

Then all quantum states can be classified into \emph{separable} or \emph{entangled} states. The \emph{set of separable states} is defined by the convex (and compact) hull of product states
\beq\label{set-separable-states}
S \;=\; \big\{\rho \,=\, \sum_{i} p_i \,\rho^A_i \otimes \rho^B_i \,|\; 0\leq p_i\leq 1\,, \sum_{i} p_i =1  \big\} \;.
\eeq
A state is called \emph{entangled} if it is not separable, i.e., $\rho_{\rm{ent}}\in S^c$ where $S^c$ denotes the complement of $S$, with $S \cup S^c = \widetilde{\cal H} \subset L(\Ha)\,$.\\

In terms of density matrices the CHSH inequality (\ref{CHSH-inequality}) can be rewritten in the following way
\beq\label{CHSH-inequality-density-matrix}
\langle \rho | \B_{\rm{CHSH}} \rangle \;=\; \rm{Tr} \,\rho \B_{\rm{CHSH}} \;\leq\; 2 \;,
\eeq
for all local states $\rho\,$, where the CHSH-Bell operator in case of qubits is expressed by
\beq\label{CHSH-Bell operator}
\B_{\rm{CHSH}} \;=\; \vec{a}\cdot\vec{\sigma}_A \otimes (\vec{b} - \vec{b}^{\,'})\cdot\vec{\sigma}_B \;+\; \vec{a}^{\,'}\cdot\vec{\sigma}_A \otimes (\vec{b} + \vec{b}^{\,'})\cdot\vec{\sigma}_B \;.
\eeq
Rewriting inequality (\ref{CHSH-inequality-density-matrix}) gives
\begin{equation}\label{CHSH-inequality-density-matrix-separable}
\langle \rho | \, 2\cdot\mathds{1} \,-\, \B_{\rm{CHSH}} \rangle \;\geq\; 0 \;.
\end{equation}
If we choose, however, the entangled Bell state $\rho^{-} = \ket{\psi^{\,-}}\bra{\psi^{\,-}}$ the inner product changes the sign
\begin{equation}\label{CHSH-inequality-density-matrix-Bellstate}
\langle \rho^{-} | \, 2\cdot\mathds{1} \,-\, \B_{\rm{CHSH}} \rangle \;<\; 0 \;.
\end{equation}

Now we can ask, is the inner product (\ref{CHSH-inequality-density-matrix-Bellstate}) negative for \emph{all} entangled states? The answer is \emph{yes} for all pure entangled states \cite{Gisin1991}, i.e., there exist measurement directions for which the CHSH inequality is violated. For mixed states, however, the situation is much more subtle. Werner \cite{werner89} discovered that a certain family of bipartite mixed states, which remained entangled, produced an outcome that admitted a local HV model for projective measurements, that is, it satisfies all possible Bell inequalities.

This feature is nicely demonstrated by the so-called  \emph{Werner states}
\begin{equation}\label{Werner-state in matrix notation}
\rho_{\,\rm{Werner}} \;=\; \alpha\, \rho^- \,+\, \frac{1-\alpha}{4}\, \mathds{1}_4 \;=\; \frac{1}{4}
\begin{pmatrix}
1-\alpha & 0        & 0        & 0 \\
0        & 1+\alpha & -2\alpha & 0 \\
0        & -2\alpha & 1+\alpha & 0 \\
0        & 0        & 0        & 1-\alpha \\
\end{pmatrix}\,,
\end{equation}
or in terms of the Bloch decomposition we have
\begin{equation}\label{Werner-state in Bloch notation}
\rho_{\,\rm{Werner}} \;\equiv\; \rho_{\alpha} \;=\; \frac{1}{4}\left(\mathds{1}\,\otimes\,\mathds{1} \,-\, \alpha\,\sigma_i\,\otimes\,\sigma_i\right) \,,
\end{equation}
with the parameter values $\alpha \in [0,1]\,$.\\

\begin{figure}
\centering
\includegraphics[width=0.45\textwidth]{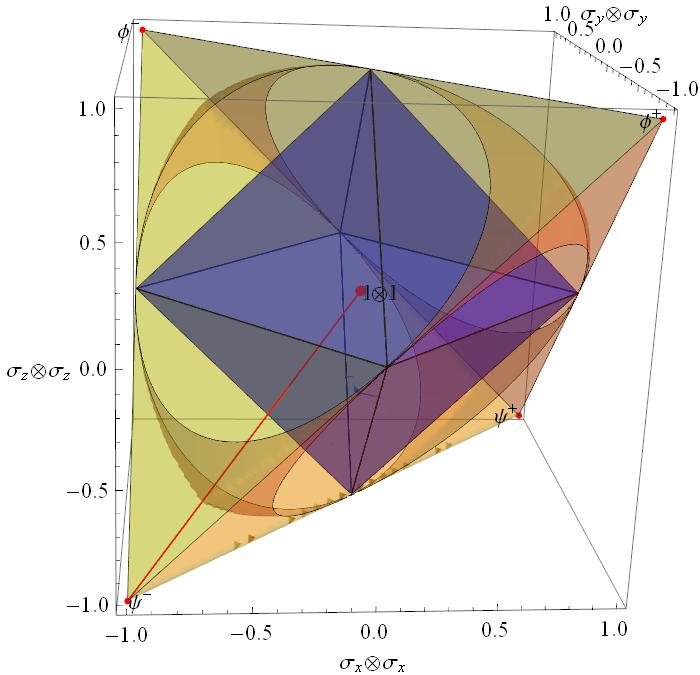}
\caption{Tetrahedron of physical states in $2 \times 2$ dimensions spanned by the four Bell states $\psi^+ , \psi^- , \phi^+ , \phi^-\,$: The separable states with the maximal mixture $\frac{1}{4}\mathds{1}_4 \,=\, \frac{1}{4}(\mathds{1}\otimes\mathds{1})$ at the origin form the blue double pyramid and the entangled states are located in the remaining tetrahedron cones. The local states satisfying a CHSH-Bell inequality lie within the parachutes (dark-yellow surfaces) containing all separable but also a lot of mixed entangled states. The Werner states (\ref{Werner-state in matrix notation}) (red line from the origin to the maximal entangled Bell state $\psi^-$) pass through all regions exhibiting the intervals of separability, locality and entanglement.}
	\label{fig:Tetrahedron-of-physical-states-Werner-line}
\end{figure}

Currently, the overall picture is the following:

The Werner states (\ref{Werner-state in matrix notation}) are separable within the bound of mixedness $\alpha \leq 1/3 = 0.33\,$ (see Theorem \ref{theorem:Peres-Horodecki}). The states admit a local HV model for all (positive-operator-valued) measurements within $\alpha \leq 5/12 = 0.42$ \cite{Barrett2002} and Werner's local HV model for projective measurements for $\alpha \leq 1/2 = 0.5$ \cite{werner89}. Focusing on projective measurements the critical value for local states can be pushed further to $\alpha \leq 1/\rm{K_G}(3) = 0.66$ (where $\rm{K_G(3)}$ is Grothendieck's constant of order 3) \cite{Acin-Gisin-Toner2006}. On the other hand, we know that Werner states (\ref{Werner-state in matrix notation}) violate the CHSH inequality for $\alpha > 1/\sqrt{2} = 0.707\,$ (see Theorem \ref{theorem:Horodecki-H-H}), thus becoming nonlocal. This bound of nonlocality can be slightly decreased to $\alpha > 0.705$ if Bell inequalities are considered that can be violated slightly stronger than the CHSH one \cite{Vertesi2008}. An illustration of these features in Hilbert-Schmidt space can be seen in Fig.~\ref{fig:Tetrahedron-of-physical-states-Werner-line}, the geometric details we will discuss in Sect.~\ref{subsec:geometry-of-quantum-states}.\\

The region of separability is determined by the so-called \emph{PPT criterion} (positive partial transposition) of Peres and the Horodecki family~\cite{peres96, horodecki96}. Given a general density matrix $\rho$ in Hilbert-Schmidt space $\widetilde{\cal H} \,=\, {\widetilde{\cal H}}_A \otimes \widetilde{{\cal H}}_B$ in its Bloch decomposition form
\begin{equation}\label{rho-general-decomposition-Bloch-form}
\rho \;=\; \frac{1}{4}\left(\,\mathds{1}\,\otimes\,\mathds{1} \,+\, r_i \,\sigma_i\,\otimes\,\mathds{1} \,+\,
u_i \,\mathds{1}\,\otimes\,\sigma_i \,+\, t_{ij} \;\sigma_i\,\otimes\,\sigma_j\right) \;,
\end{equation}
then a \emph{partial transposition} is defined by the operator $T$ acting in a subspace ${\widetilde{\cal H}}_A$ or ${\widetilde{\cal H}}_B$ and transposing there the off-diagonal elements of the Pauli matrices: $T\,(\sigma^i)_{kl} \,=\, (\sigma^i)_{lk}\,$.\\

\begin{theorem}[PPT-Criterion~\cite{peres96, horodecki96}]\ \

If $\quad (\mathds{1}_A \otimes T_B)\,\rho \;\geq\; 0 \;\;$ or $\;\; (T_A\otimes\mathds{1}_B)\,\rho \;\geq\; 0$ $\quad \Longleftrightarrow \quad$ $\rho$ separable.\\
\label{theorem:Peres-Horodecki}
\end{theorem}

Theorem \ref{theorem:Peres-Horodecki} holds only in $2 \times 2$ and $2 \times 3$ dimensions. We say a \emph{PPT state} is a state that remains positive under partial transposition. In higher dimensions the PPT criterion is only necessary but not sufficient for separability. If, however, at least one of the eigenvalues of the partially transposed matrix is negative we call it a \emph{NPT state}, and it has to be entangled. But there exist entangled states that remain positive semidefinite, \emph{PPT entangled states}, these are called \emph{bound entangled states}, since they cannot be distilled to a maximally entangled state (see the literature~\cite{horodecki98, HHHH07, bertlmann-krammer-AnnPhys09, bertlmann-krammer-PRA-78-08, bertlmann-krammer-PRA-77-08, baumgartner-hiesmayr-narnhofer06, baumgartner-hiesmayr-narnhofer07, baumgartner-hiesmayr-narnhofer08}).\\

In our case of the Werner states (\ref{Werner-state in matrix notation}) partial transposition acts like
\begin{equation}
(\mathds{1}_A \otimes T_B)\,\rho_{\,\rm{Werner}} \;=\; \frac{1}{4}
\begin{pmatrix}
1-\alpha & 0        & 0        & -2\alpha \\
0        & 1+\alpha & 0        & 0 \\
0        & 0        & 1+\alpha & 0 \\
-2\alpha & 0        & 0        & 1-\alpha \\
\end{pmatrix}\,,
\label{partial-transposition-Werner-state}
\end{equation}
which has as eigenvalues $\lambda_{1,2,3} = \frac{1 + \alpha}{4}$ and $\lambda_{4} = \frac{1 - 3\alpha}{4}$. Therefore the states are separable for $\alpha \leq \frac{1}{3}$ and entangled for $\alpha > \frac{1}{3}\,$.\\

In order to find the states violating a Bell inequality an other theorem of the Horodecki family~\cite{horodecki95} is very powerful since we do not have to check all measurement directions $\vec{a}$ and $\vec{b}\,$.

\begin{theorem}[Maximal violation of a Bell inequality \cite{horodecki95}]\ \
Given a general $2 \times 2$ dimensional density matrix in Bloch form $\rho \,=\, \frac{1}{4}\left(\,\mathds{1}\,\otimes\,\mathds{1} + r_i \,\sigma_i\,\otimes\,\mathds{1} +
u_i \,\mathds{1}\,\otimes\,\sigma_i + t_{ij} \;\sigma_i\,\otimes\,\sigma_j\right) \,$ and the Bell operator
\begin{equation}\label{Bell operator}
\frac{1}{2}\,\B_{\rm{CHSH}} \;=\; \frac{1}{2}\,\big( \vec{a}\cdot\vec{\sigma}_A \otimes (\vec{b} - \vec{b}^{\,'})\cdot\vec{\sigma}_B \;+\; \vec{a}^{\,'}\cdot\vec{\sigma}_A \otimes (\vec{b} + \vec{b}^{\,'})\cdot\vec{\sigma}_B\big) \,,
\end{equation}
then the maximal violation of the Bell inequality $B^{\rm{max}} \,=\, \frac{1}{2}\,\rm{max}_{\B}\,\rm{Tr}\,\rho\,\B_{\rm{CHSH}}\,$ is given by
\begin{equation}\label{Bell operator maximal value}
B^{\rm{max}} \;=\; \sqrt{t^2_1 + t^2_2} \;>\; 1 \,,
\end{equation}
where $t^2_1, t^2_2\,$ denote the two larger eigenvalues of the matrices product $(t_{ij})^T (t_{ij})\,$.\\
\label{theorem:Horodecki-H-H}
\end{theorem}

In case of the Werner states we can read off the coefficient matrix directly from the Bloch decomposition (\ref{Werner-state in Bloch notation})
\begin{equation}
(t_{ij}) \;=\; -
\begin{pmatrix}
\alpha   & 0      & 0        \\
0        & \alpha & 0        \\
0        & 0      & \alpha   \\
\end{pmatrix}\,,
\label{coefficient-matrix-Werner-state}
\end{equation}
which yields the maximal violation of the CHSH inequality by $B^{\rm{max}} \;=\; \sqrt{2 \alpha^2} \;>\; 1 \,$. Thus, for all $\alpha > 1/\sqrt{2}$ the CHSH-Bell inequality is violated.

\subsection{Entanglement Witness}

From the above analysis we see that a Bell operator given by expression (\ref{CHSH-Bell operator}) is not appropriate to find all entangled states. In order to locate entanglement a different operator has to be constructed. Quite generally, entanglement can be `detected' by an Hermitian operator, the so-called \emph{entanglement witness} $A$, that detects the entanglement of a state $\rho_{\rm ent}$ via the \emph{entanglement witness inequalities} (for detailed literature, see Refs.~\cite{horodecki96, HHHH07, terhal00, bertlmann-narnhofer-thirring02, bruss02, bruss-cirac-horodecki-etal02, guehne-toth-PhysRep2009, krammer-DA-2005, gabriel-DA-2009}).\\

\begin{theorem}[Entanglement Witness Theorem~\cite{horodecki96, terhal00, bertlmann-narnhofer-thirring02}]\ \

A state $\rho_{\rm ent}$ is entangled if and only if there is a Hermitian operator $A$ -- the entanglement witness -- such that
\begin{eqnarray} \label{def-entwit}
    \left\langle \rho_{\rm ent}|A \right\rangle \;=\; \textnormal{Tr}\, \rho_{\rm ent} A
    & \;<\; & 0 \,,\nonumber\\
    \left\langle \rho|A \right\rangle = \textnormal{Tr}\, \rho A & \;\geq\; & 0 \qquad
    \forall \rho \in S \,,
\end{eqnarray}

where $S$ denotes the set of all separable states.
\label{theorem:entanglement-witness-theorem}
\end{theorem}

\begin{figure}
\centering
\includegraphics[width=0.3\textwidth]{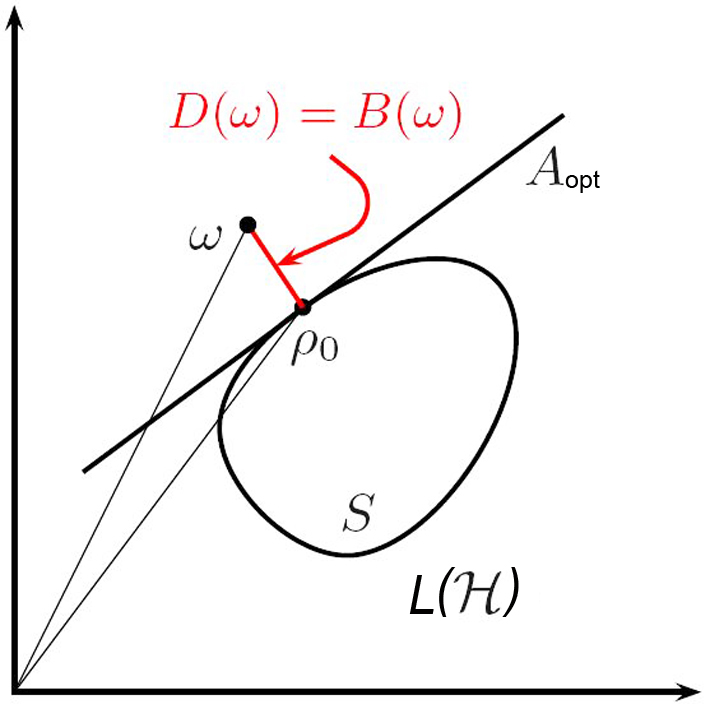}
\caption{Example of an entangled state $\rho_{\rm ent} = \omega$ and the set of separable states $S\,$. The tangent plane $A_{\rm{max}}$ represents the optimal entanglement witness $A_{\rm{opt}}$ and $\rho_0$ the nearest separable state to $\omega$. Theorem~\ref{theorem:BNT-theorem} is illustrated; it states that the maximal violation $B(\omega)$ of the entanglement witness inequality (\ref{def-entwit}) is equal to the minimal Hilbert-Schmidt distance of the entangled state $\omega$ to the set $S$ of separable states, which represents a measure of entanglement.}
\label{fig:illustration-of-BNT-theorem}
\end{figure}

An entanglement witness is \emph{optimal}, denoted by $A_{\rm opt}\,$, if apart from Eq.~(\ref{def-entwit}) there exists a separable state $\rho_0 \in S$ such that
\begin{equation}
    \left\langle \rho_0 |A_{\rm opt} \right\rangle \;=\; 0 \,.
\end{equation}
The operator $A_{\rm opt}$ defines a tangent plane to the convex set of separable states $S$ (\ref{set-separable-states}) (see Fig.~\ref{fig:illustration-of-BNT-theorem}). Such an $A_{\rm opt}$ always exists due to the Hahn-Banach Theorem and the convexity of $S\,$.\\

On the other hand, with help of the Hilbert-Schmidt norm we can define the \emph{Hilbert-Schmidt distance} between two arbitrary states $\rho_1$ and $\rho_2$
\begin{equation}
    d_{\rm{HS}}(\rho_1,\rho_2) \;=\; \| \rho_1 - \rho_2 \| \;=\; \sqrt{<\rho_1 - \rho_2 | \rho_1 - \rho_2>} \;=\; \sqrt{\rm{Tr} \,(\rho_1 - \rho_2)^\dagger (\rho_1 - \rho_2)}  \;.
\end{equation}
We can view the minimal distance of an entangled state $\rho_{\rm ent}$ to the set of separable states, the \emph{Hilbert-Schmidt measure}
\begin{equation} \label{def-HSmeasure}
    D(\rho_{\rm{ent}}) \;:=\; \min_{\rho \in S} \left\| \rho \,-\, \rho_{\rm{ent}} \right\|
    \;=\; \left\| \rho_0 \,-\, \rho_{\rm{ent}} \right\| \,,
\end{equation}
where $\rho_0$ denotes the nearest separable state, as a \emph{measure for entanglement}.\\

There is an interesting connection between the Hilbert-Schmidt measure and the entanglement witness inequality. Let us rewrite entanglement witness inequality (\ref{def-entwit})
\beq\label{EWIrewritten}
\left\langle \rho|A \right\rangle \,-\, \left\langle \rho_{\rm ent}|A \right\rangle \;\geq\; 0 \qquad \forall \rho \in S \,,
\eeq
and define the maximal violation of inequality (\ref{EWIrewritten}) as follows ($\rho$ and $A$ are still free at our disposal):

\begin{definition}[Maximal violation of the entanglement witness inequality~\cite{bertlmann-narnhofer-thirring02}]\ \
\begin{eqnarray} \label{max-violation-EWI}
   B(\rho_{\rm ent}) \;=\; \max_{A} \,\big( \min_{\rho \in S} \left\langle \rho|A \right\rangle \,-\, \left\langle \rho_{\rm ent}|A \right\rangle  \big) \,.
\end{eqnarray}
\label{definiton:max-violation-of-EWI}
\end{definition}
The minimum is taken over all separable states and maximum over all possible entanglement witnesses $A\,$, suitably normalized. Then there holds the following theorem~\cite{bertlmann-narnhofer-thirring02, krammer-DA-2005, gabriel-DA-2009}:

\begin{theorem}[Bertlmann-Narnhofer-Thirring Theorem~\cite{bertlmann-narnhofer-thirring02}]\ \
\begin{eqnarray} \label{B-equals-D}
    a)&& B(\rho_{\rm ent}) \;=\; D(\rho_{\rm{ent}}) \,,
\end{eqnarray}
    b) The maximal violation of the entanglement witness inequality is achieved when $\rho \rightarrow \rho_0$ and $A \rightarrow A_{\rm opt}\,$, then the optimal entanglement witness is given by \\
\begin{eqnarray} \label{Aopt-explicit-expression}
    A_{\rm opt} \;=\; \frac{\rho_0 \,-\, \rho_{\rm ent} \,-\, \left\langle \rho_0|\rho_0 \,-\, \rho_{\rm ent} \right\rangle \mathds{1}}{\left\| \rho_0 \,-\, \rho_{\rm ent} \right\|} \;.
\end{eqnarray}
\label{theorem:BNT-theorem}
\end{theorem}

The maximal violation (\ref{max-violation-EWI}) of the entanglement witness inequality is equal to the Hilbert-Schmidt measure (\ref{def-HSmeasure}), which is a measure of entanglement. This is a remarkable result (see Fig.~\ref{fig:illustration-of-BNT-theorem}). The optimal entanglement witness is given explicitly by expression (\ref{Aopt-explicit-expression}). However, we have to know the nearest separable state $\rho_0\,$, which is easy to find in low dimensions, but not in higher ones. Nevertheless, there exist approximation procedures to approach $\rho_0$ (see Ref.~\cite{bertlmann-krammer-AnnPhys09}).\\

For example, in case of Alice and Bob the Werner states are given by $\rho_{\alpha}$ (\ref{Werner-state in Bloch notation}), and the Bell state $\rho^{-} = \ket{\psi^{\,-}}\bra{\psi^{\,-}}$ by choosing the parameter value $\alpha = 1\,$, i.e., $\rho^{-} = \rho_{\alpha = 1}\,$.

The nearest separable state is easily found
\beq\label{nearest-sep-state-Alice-Bob}
\rho_0 \;=\; \frac{1}{4}\left(\mathds{1}\,\otimes\,\mathds{1} \,-\, \frac{1}{3}\,\sigma_i\,\otimes\,\sigma_i\right) \,,
\eeq
yielding the Hilbert-Schmidt measure
\beq\label{HSmeasure-for-Alice-Bob}
D(\rho_{\alpha}) \;=\; \left\| \rho_0 \,-\, \rho_{\alpha} \right\| \;=\; \frac{\sqrt{3}}{2}\,(\alpha \,-\, \frac{1}{3}) \,.
\eeq
The optimal entanglement witness we calculate from expression (\ref{Aopt-explicit-expression})
\begin{equation}\label{Aopt-for-Alice-Bob}
A_{\rm opt} \;=\; \frac{1}{2\sqrt{3}}\left(\mathds{1}\,\otimes\,\mathds{1} \,+\, \sigma_i\,\otimes\,\sigma_i\right) \,,
\end{equation}
and the maximal violation of the entanglement witness inequality from Eq.~(\ref{max-violation-EWI})
\beq\label{max-violat-of-EWI-for-Alice-Bob}
B(\rho_{\alpha}) \;=\; -\, \left\langle \rho_{\alpha}|A_{\rm opt} \right\rangle \;=\; \frac{\sqrt{3}}{2}\,(\alpha \,-\, \frac{1}{3}) \,.
\eeq
Clearly, both results (\ref{max-violat-of-EWI-for-Alice-Bob}) and (\ref{HSmeasure-for-Alice-Bob}) coincide as required by Theorem~\ref{theorem:BNT-theorem}.

\subsection{Geometry of Quantum States}\label{subsec:geometry-of-quantum-states}

The quantum states for a two-qubit system, the case of Alice and Bob, have very nice geometric features in the Hilbert-Schmidt space, more precisely, in the spin-spin space. Quite generally a quantum state can be decomposed as in Eq.~(\ref{rho-general-decomposition-Bloch-form}), where the last term, the spin-spin term, is the important one to characterize entanglement. If we parameterize the spin-spin space by
\beq\label{Bell-states-parametrisation-Alice-Bob}
\rho \;=\; \frac{1}{4}\left(\mathds{1}\,\otimes\,\mathds{1} \,+\, \sum_i\,c_i\,\sigma_i\,\otimes\,\sigma_i \right) \,,
\eeq
the Bell states have the coefficients $c_i = \pm 1$.

Due to the positivity of the density matrix the four Bell states $\psi^{-}, \psi^{+}, \phi^{-}, \phi^{+}$ set up a simplex, a tetrahedron, in this spin-spin space \cite{bertlmann-narnhofer-thirring02, vollbrecht-werner-PRA00, horodecki-R-M96} (see Fig.~\ref{fig:Tetrahedron-of-physical-states-Werner-line}). The separable states, given by the PPT criterion (Theorem~\ref{theorem:Peres-Horodecki}), form an octahedron which lies inside, and the maximal mixed state $\frac{1}{4}\mathds{1}_4 \,=\, \frac{1}{4}(\mathds{1}\otimes\mathds{1})$ is placed at origin. The entangled states are located in the remaining cones. The local states, on the other hand, satisfying a CHSH-Bell inequality lie within the parachutes, the dark-yellow surfaces in Fig.~\ref{fig:Tetrahedron-of-physical-states-Werner-line}.
They are determined by Theorem~\ref{theorem:Horodecki-H-H} and contain all separable but also a large amount of mixed entangled states. The Werner states (\ref{Werner-state in matrix notation}), the red line in Fig.~\ref{fig:Tetrahedron-of-physical-states-Werner-line} from the origin to the maximal entangled Bell state $\psi^-\,$, show nicely how the states change from maximal mixed and separable to local, mixed entangled, and finally to nonlocal states, ending at $\psi^{-}$ which is pure and maximal entangled.\\

\begin{figure}
\centering
\includegraphics[width=0.45\textwidth]{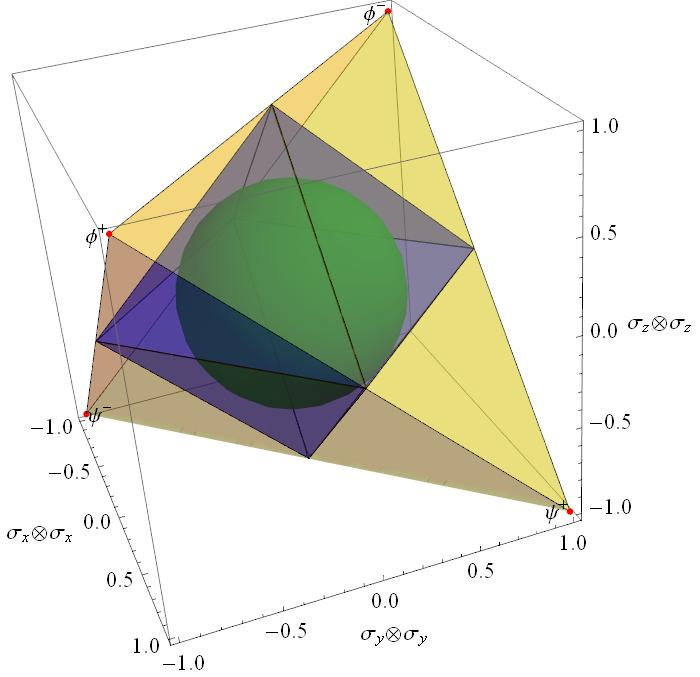}
\caption{Tetrahedron of physical states in $2 \times 2$ dimensions spanned by the four Bell states $\psi^+ , \psi^- , \phi^+ , \phi^-\,$: The separable states form the blue double pyramid and the entangled states are located in the remaining tetrahedron cones. The unitary invariant maximal ball (shaded in green) of absolutely separable states is placed within the double pyramid and the maximal mixture $\frac{1}{4}\mathds{1}_4$ is at the origin.}
\label{fig:Tetrahedron-of-physical-states-absolutely-separable-ball}
\end{figure}

Finally, I want to make a remark about the entanglement and separability of the quantum states. When talking about entanglement it is important to refer to the chosen factorization of the algebra of a density matrix. Depending on the considered factorization, a quantum state appears either entangled or separable \cite{TBKN}. For pure states we always can switch unitarily between separability and entanglement, however, for mixed states a minimal amount of mixedness is needed. It is interesting to search for this bound of mixedness. Within this bound the states are separable with respect to all possible factorizations, i.e., they remain separable for all unitary transformations. Such states are called \emph{absolutely separable states} \cite{TBKN, kus-zyczkowski, zyczkowski-bengtsson, bengtsson-zyczkowski-book}, the maximal mixed state $\frac{1}{4}\mathds{1}_4$ represents the prototype. When considering the \emph{maximal ball} around $\frac{1}{4}\mathds{1}_4$ of constant mixedness, which can be inscribed into the set of mixed states for a bipartite system then all states belonging to that ball are not only separable but also absolutely separable (see the green shaded ball in Fig.~\ref{fig:Tetrahedron-of-physical-states-absolutely-separable-ball}).

\section{Bell Experiments -- a Historical View}\label{sec:Bell experiments}

\subsection{First Generation Experiments of the Seventies}

It is interesting, in their quantum debate neither Einstein in his EPR paper~\cite{EPR} nor Bohr in his reply~\cite{Bohr} to EPR proposed an experiment that could be performed, the discussion was pure theoretically or philosophically. The first step towards an experimental verification was laid by Bohm in his spin version~\cite{Bohm-Aharanov-EPRspin1957} of the EPR paradox. John Bell~\cite{Bell-Physics1964} developed it further and discovered the inequality that all local realistic theories had to satisfy. But also Bell's work was yet not quite adapted to experiment.

In the late sixties, John Clauser, a young graduate student from Columbia University, read Bohm's~\cite{Bohm1952} and de Broglie's~\cite{deBroglie1960, deBroglie1963, deBroglie1964} work about an alternative view of quantum mechanics. While he had difficulties in understanding the Copenhagen interpretation of QM, he found the works of these two critics quite sensible. When studying next Bell's inequality paper~\cite{Bell-Physics1964} that contained a bound for all hidden variable theories, he was astounded by its result. As a true experimentalist he wanted to see the experimental evidence for it. So he planed to do the experiment.\\

However, experiments of this type were not appreciated at that time, it was the ``dark era'' of the foundations of quantum mechanics. Pauli's opinion was often cited~\cite{Pauli-letter-to-Born1954}:

\emph{``One should no more rack one's brain about the problem of whether something one cannot know anything about exists at all, than about the ancient question of how many angels are able to sit on the point of a needle.''}\\

When Clauser had an appointment with Richard Feynman at Caltech to discuss an experimental EPR configuration for testing the predictions of QM, he immediately threw him out of his office saying~\cite{Clauser2002}:

\emph{``Well, when you have found an error in quantum-theory's experimental predictions, come back then, and we can discuss your problem with it.''}\\

But Clauser remained stubborn, he belonged to the revolting generation, and prepared the experiment. He sent an Abstract to the Spring Meeting of the American Physical Society proposing the experiment~\cite{Clauser1969}. Soon afterwards, Abner Shimony called and told him that he and his student Michael Horne had the same ideas. So they joined and wrote together with Richard Holt, a PhD student of Francis Pipkin from Harvard, the famous CHSH paper~\cite{CHSH}, where they proposed an inequality that was well adapted to experiments.\\

Clauser carried out the experiment in 1972 together with Stuart Freedman~\cite{Clauser-Freedman1972}, a graduate student at Berkeley, who received his PhD with this experiment. As pointed out in the CHSH paper~\cite{CHSH}, pairs of photons emitted in an atomic radiative cascade would be suitable for a Bell inequality test. Clauser and Freedman chose Calcium atoms pumped by lasers, where the excited atoms emitted the desired photon pairs. The signals were very weak at that time, a measurement lasted for about 200 hours. For comparison with theory a very practical inequality was used, which was derived by Freedman~\cite{Freedman1972}. The outcome of the experiment is well known, they obtained a clear violation of the Bell inequality very much in accordance with QM.\\

At the same time, in competition to Clauser's experiment, Pipkin and Holt performed a similar experiment at Harvard using a radiative cascade in mercury. The outcome was just contrary, the results agreed with HVT and disagreed with QM~\cite{Holt-Pipkin1974}. Knowing about Clauser's result they withdrew the paper from publication. Clauser repeated the Holt-Pipkin experiment with the result that it violated the Bell inequality significantly~\cite{Clauser1976}.\\

Independently in 1976 at Houston, Edward Fry and his student Randall Thompson set up an experiment by using mercury atoms. As in Clauser's experiment the correlated photons were produced in a radiative cascade from by lasers excited atomic levels. Due to the much better signals with improved lasers they could collect enough data already in 80 minutes. The result was in excellent agreement with QM, the Bell inequality was violated by 4 standard deviations~\cite{Fry-Thompson1976}.\\

Different kind of experiments, positron annihilation experiments~\cite{Kasday-etal1970, Faraci-etal1974, Wilson-etal1976, Bruno-etal1977, Bertolini-etal1981} and proton-proton scattering~\cite{Lamehi-Rachti-Mittig1976} pointed into the same direction.\\

Thus at that time, it was already convincing that hidden variable theories did not work but quantum mechanics was correct. However, the experiments were not perfect yet, the analyzers were static, only a small amount of photon pairs were registered, etc. There still existed several loopholes, the detection efficiency or fair sampling loophole, and the communication absence or freedom of choice loophole, just to mention some important ones. To close these loopholes was the challenge of the future experiments.

\subsection{Second Generation Experiments of the Eighties}\label{second-generation-experiments}

In the late 1970s and beginning of the 1980s, the general atmosphere in the physics community was still such: \emph{``Quantum mechanics works very well, so don't worry!''} Nevertheless, Alain Aspect, when reading Bell's inequality paper~\cite{Bell-Physics1964}, was so strongly impressed that he immediately decided to do his \emph{``th\`ese d'\' etat''} on this fascinating topic. He visited John Bell at CERN to discuss his proposal. John's first question was, as Aspect told me, \emph{``Do you have a permanent position?''} Only after Aspect's positive answer the discussion could begin. Aspect's goal was to include variable analyzers.

Aspect and his collaborators performed a whole series of experiments~\cite{Aspect1976, Aspect-etal1981, Aspect-etal1982, Aspect-Dalibard-Roger1982, Aspect-etal1985} with an improved design and approached step by step the `ideal' setup configuration. As Clauser, they chose a radiative cascade in calcium that emitted photon pairs in the Bell state $\ket{\phi^{\,+}}\,$. For comparison with theory the Clauser-Horne inequality (\ref{Clauser-Horne-inequality}) was used, which was significantly violated in each experiment.

In the final time-flip experiment~\cite{Aspect-Dalibard-Roger1982} together with Jean Dalibard and G\' erard Roger a clever acoustic-optical switch mechanism was incorporated. It worked such that the switching time between the polarizers, as well as the lifetime of the photon cascade, was much smaller than the time of flight of the photon pair from the source to the analyzers. That implied a space-like separation of the event intervals. However, the time flipping mechanism was still not ideal, i.e., truly random, but \emph{``quasi-periodic''}, as they called it. The mean for two runs which lasted about 2 hours yielded the result $S_{\rm CH}^{\rm exp} \,=\, 0.101 \pm 0.020$ in very good agreement with the quantum mechanical prediction $S_{\rm CH}^{\rm QM} \,=\, 0.113 \pm 0.005$ that had been adapted for the experiment (recall the ideal value is $S_{\rm CH}^{\rm QM} \,=\, 0.207$ (\ref{Clauser-Horne-inequality-max-violation})).\\

Some time later, a different experiment was carried out by Hans Kleinpoppen and his group~\cite{Kleinpoppen-etal1985}. They used metastable atomic deuterium for emitting spontaneously two correlated photons which were measured. The data violated a Bell inequality in Freedman's form~\cite{Freedman1972} and were in agreement with QM.\\

I remember that the time-flip experiment of Aspect received much attention in the physics community and also in popular science, and Alain Aspect was the best apologist. In my opinion, it caused a turning point, the physics community began to realize that there was something essential in it. The research started and flourished into a new direction, into what is called nowadays quantum information and quantum communication.

\subsection{Third Generation Experiments of the Nineties and Beyond}\label{third-generation-experiments}

In the 1990s, the spirit towards foundations in quantum mechanics totally changed since quantum information gained increasing interest, Bell inequalities and quantum entanglement were the basis.

\begin{figure}
\centering
\includegraphics[width=0.4\textwidth]{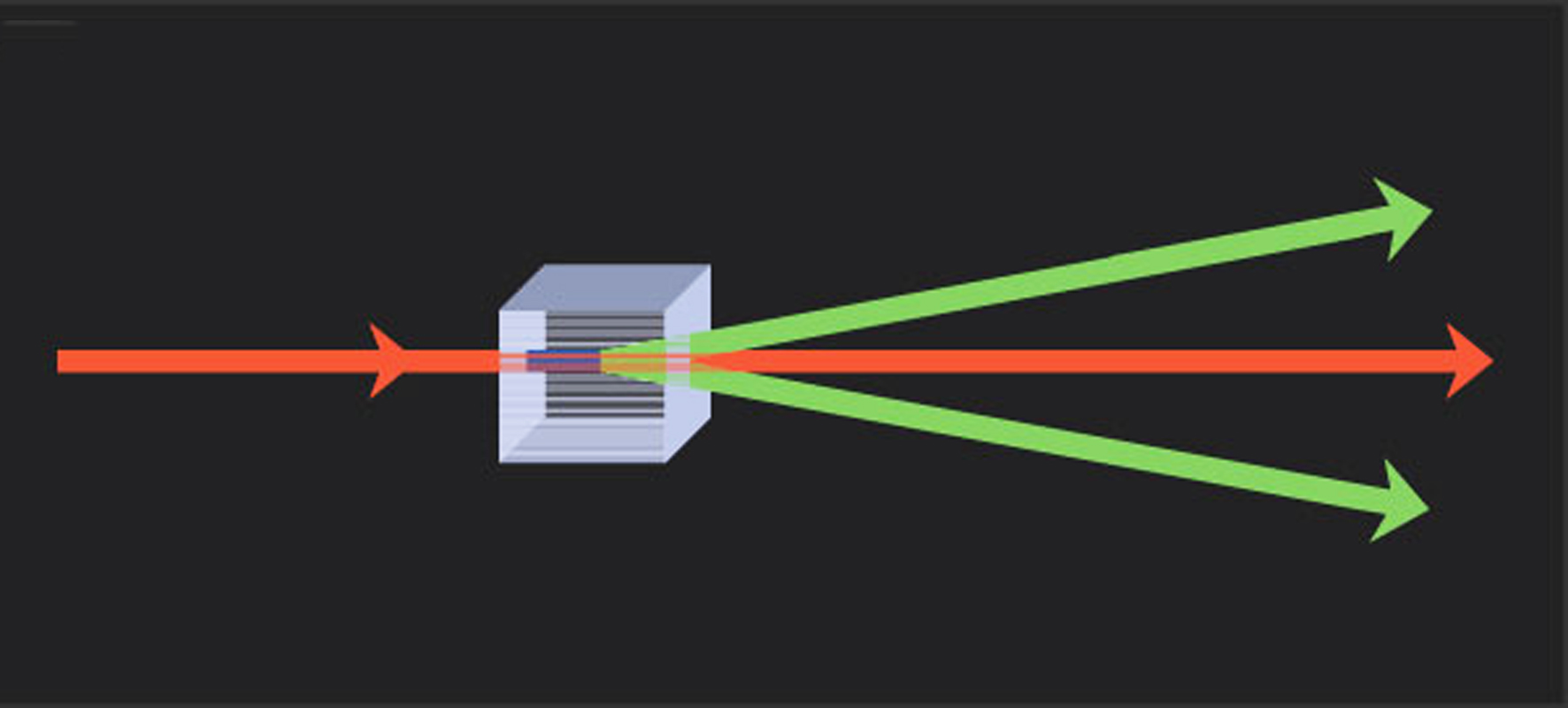}\\

\centering
\includegraphics[width=0.4\textwidth]{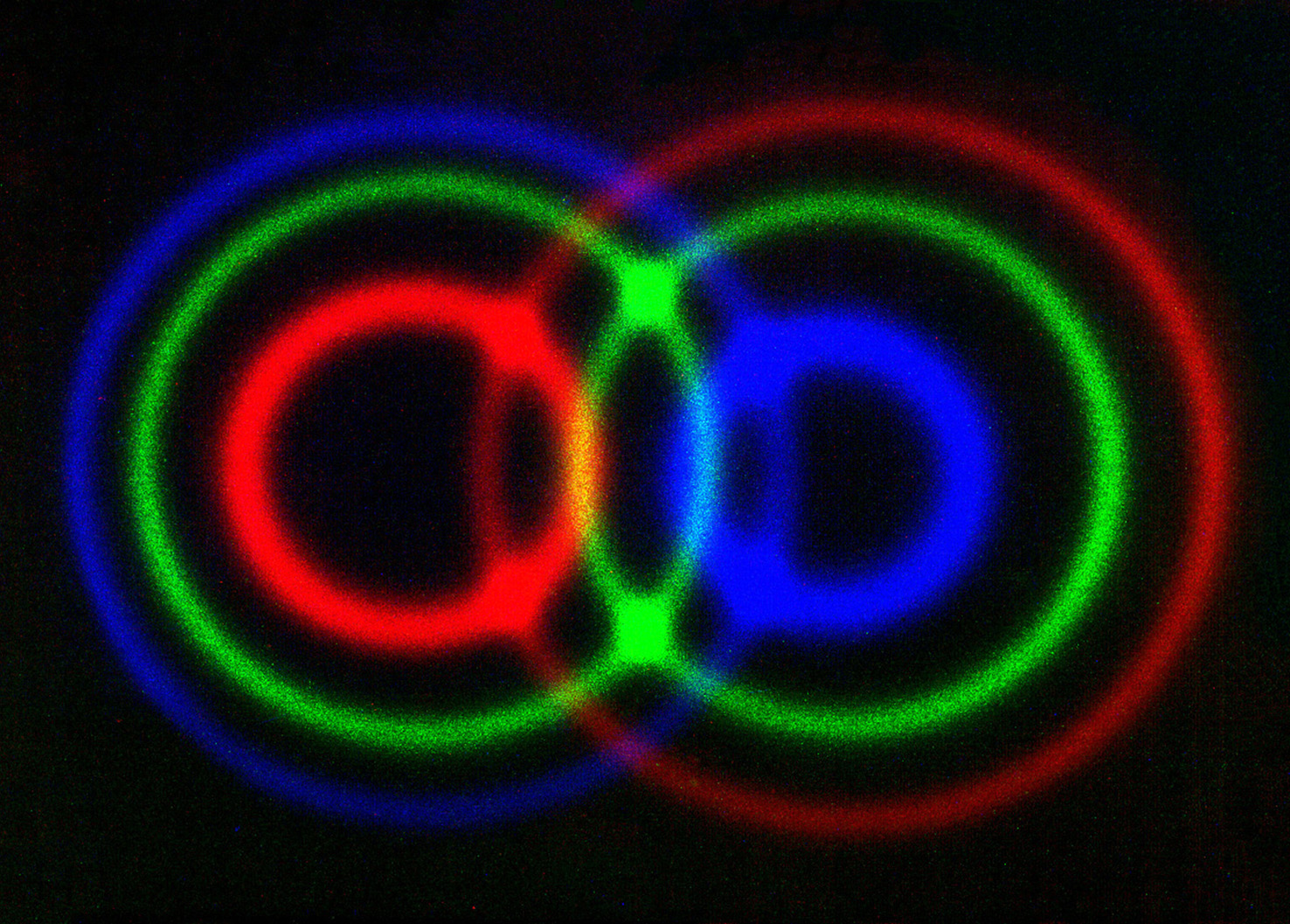}
\caption{Above: A BBO crystal is pumped by a laser, the outgoing photons are vertically and horizontally polarized on two different cones. Foto: \copyright IQOQI, Vienna. Below: A cut through the two different cones; in the overlap regions the photons are entangled. Foto: \copyright Faculty of Physics, University of Vienna, Foto taken by Paul Kwiat and Michael Reck.}
	\label{fig:parametric-downconversion-entangled-photons}
\end{figure}

Meanwhile, the technical facilities improved considerably too, the electronics and the lasers. Most important was the invention of a new source for creating two entangled photons. That was spontaneous parametric down conversion, where a nonlinear crystal was pumped with a laser and the pump photon was converted into two photons that propagated vertically and horizontally polarized on two different cones. In the overlap region they were entangled (see Fig.~\ref{fig:parametric-downconversion-entangled-photons}).

\begin{figure}
\centering
\includegraphics[width=0.8\textwidth]{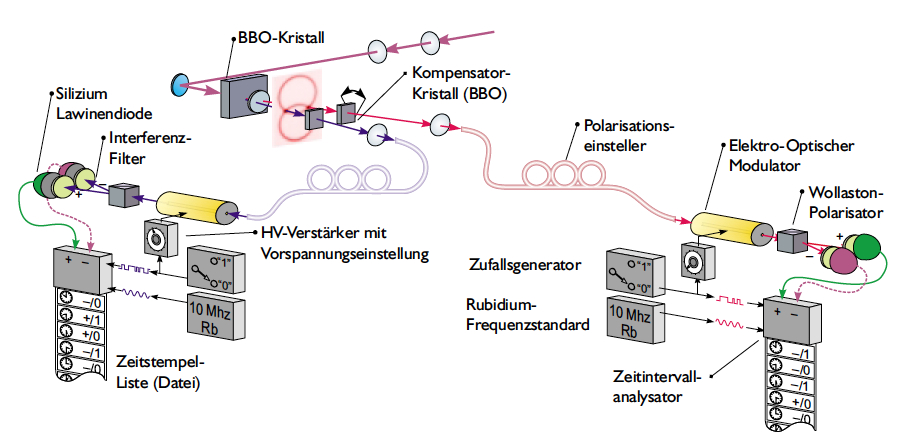}
\caption{The timing experiment of Weihs et al.~\cite{Weihs-Zeilinger-Bell-experiment1998}. The EPR source is a so-called BBO crystal pumped by a laser, the outgoing photons are vertically and horizontally polarized on two different cones and in the overlap region they are entangled. This entangled photons are led separately via optical fibres to the measurement stations Alice and Bob. During the photon propagation the orientations of polarizations are changed by an electro-optic modulator which is driven by a truly random number generator, on each side. In this way the strict Einstein locality condition---no mutual influence between the two observers Alice and Bob---is achieved in the experiment. The figure is taken from Ref.~\cite{Weihs-thesis-Bell-experiment1998}, \copyright Gregor Weihs.}
	\label{fig:Weihs-Bell-experiment}
\end{figure}

Such an EPR source was used by Anton Zeilinger and his group, when they performed their Bell experiment in 1998~\cite{Weihs-Zeilinger-Bell-experiment1998}. Zeilinger's student Gregor Weihs obtained his PhD with this experiment~\cite{Weihs-thesis-Bell-experiment1998} (see Fig.~\ref{fig:Weihs-Bell-experiment}). Their challenging goal was to construct an ultra-fast and truly random setting of the analyzers at each side of Alice and Bob, such that strict Einstein locality---no mutual influence between the two observers Alice and Bob---was achieved in the experiment. The data were compared with the CHSH inequality (\ref{CHSH-inequality}) and the experimental result was: $S_{\rm{CHSH}}^{\rm{exp}} \,=\, 2.73 \pm 0.02\,$, which corresponded to a violation of the inequality of $30$ standard deviations. Due to this high efficiency photon source the measurement could be performed already in $3-4$ minutes. It was \emph{the} experiment that truly included the vital time factor, John Bell insisted upon so strongly.\\

About the same time other groups investigated energy correlated photon pairs to test Bell inequalities~\cite{Brendel-etal1992, Tapster-etal1994}. A record was set by the group of Nicolas Gisin~\cite{Tittel-Gisin-etal1998}, by using energy-time entangled photon pairs in optical fibres. They managed to separate their observers Alice and Bob by more than 10 km and could show that this distance had practically no effect on the entanglement of the photons. The investigated Bell inequalities had been violated by 16 standard deviations.

Fascinating experiments on quantum teleportation~\cite{Bouwmeester-Zeilinger-etal1997, Ma-Zeilinger-etal-teleportation-144km-Nature2012} and quantum cryptography~\cite{Jennewein-Zeilinger-etal-QuCrypto2000, Gisin-etal-QuCrypto2002} followed.

Then a race started in achieving records of entanglement based long distance quantum communication. The vision was to install a global network, in particular via satellites or the International Space Station, that provided an access to secure communication via quantum cryptography at any location.

It was again Zeilinger's group that pushed the distance limits further and further. Firstly, in an open air experiment in the City of Vienna over a distance of $7.8$ km the group~\cite{Resch-Blauensteiner-Zeilinger-etal-freespace-City2005} could violate a CHSH inequality (\ref{CHSH-inequality}) by more than $13$ standard deviations. Secondly, this is presently the world record, the group~\cite{Ursin-Blauensteiner-Zeilinger-etal-Tenerife2007} successfully carried out an open air Bell experiment over $144$ km between the two Canary Islands La Palma and Tenerife.\\

In search of closing loopholes a recent Bell experiment of the group~\cite{Giustina-Zeilinger-fairsampling-Nature2013} closed the fair-sampling loophole, i.e., their results of violating an inequality \` a la Eberhard~\cite{Eberhard-inequ-1993} were valid without assuming that the sample of measured photons accurately represented the entire ensemble.

Another loophole, the detection efficiency loophole, could be closed with ion traps. Working with ions the group Rowe et al.~\cite{Rowe-Wineland-Bell-ions-Nature2001} tested a Bell inequality with perfect detection efficiency.\\

Finally, I also want to refer to Bell inequality tests of the group of Helmut Rauch~\cite{Hasegawa-etal-KochenSpecker2003, Hasegawa-etal-KochenSpecker2010, Hasegawa-etal-contextuality2009}. These neutron interferometer experiments were of particular interest since in this case the quantum correlations were explored in the degrees of freedom of a single particle, the neutron. Physically, it meant that rather contextuality was tested than nonlocality in space.\\

It is quite interesting and amusing to see the development of Bell experiments in the history of time. Beginning in the 1970s, where one had to overcome huge technical difficulties and the enormous resistance of the physics community, the development ended in the 2010s in such a way that a Bell experiment belonged already to the standard educational program \emph{``Laboratory Quantum Optics''} for the students at the Faculty of Physics of the University of Vienna. It would have been nice if John Bell could have seen that!

\subsection{Bell Experiments in Particle Physics}

The quantum correlations discovered in photon physics had been also found in particle physics. The particle--antiparticle systems that were generated in the huge particle accelerators were already entangled due to conservations laws. The \emph{strangeness system} $K^0 \bar{K}^0$, produced at the $\Phi$ resonance in the $e^{+} e^{-}$ machine DA$\Phi$NE at Frascati, or the \emph{beauty system} $B^0 \bar{B}^0$ produced at the $\Upsilon$(4S) resonance at KEK, Japan, were typical examples.

The difference to the photon systems, discussed so far, was that particle systems had entirely different and additional properties, which the photons did not have. First of all, the investigated particles were very massive, they decayed into other particles, they oscillated between their flavour content, i.e., between their particle and antiparticle nature, and they could regenerate. In addition, they possessed internal symmetries, like the $CP$ symmetry (charge conjugation and parity), which turned out to be essential. For these reasons, I think, that it was, and still is, of great importance to investigate such systems, particularly, with regard to the EPR-Bell quantum correlations.

The typical feature of these particle systems, e.g., of a kaon--antikaon system, was that the joint expectation value of a measurement at Alice's and Bob's detectors depended on both, on the flavour content, which corresponded to a quasi-spin property, and on the time of the measurement, once the system was created. Correspondingly, also a Bell inequality depended on both~\cite{Bertlmann-Hiesmayr2001} (see Refs.~\cite{Bertlmann-LectureNotes2006, CarlaSchuler2014}, for an overview in this field).\\

Therefore we might choose in a Bell inequality:

a) \emph{varying the flavour (quasi-spin) or fixing the time},

b) \emph{fixing the flavour (quasi-spin) or varying the time}.\\

Quite generally, the experimental test of Bell inequalities in particle physics was much more intricate than in photon physics. Active measurements had to be carried out, however, they were difficult to achieve. Usually, the measurements were passive since they happened through the decays of the particles~\cite{Bertlmann-Bramon-Garbarino-Hiesmayr2004}.\\

Let me mention two important cases, further quantum proposals can be found in Ref.~\cite{Amelino-Camelia-KLOEcoll2010}:

\emph{Case} a)
By varying the flavour content in the particle--antiparticle system a Wigner-type inequality, like Eq.~(\ref{Wigners-inequality}), could be established for the kaon system, which related the violation of the Bell inequality to $CP$ violation~\cite{Bertlmann-Grimus-Hiesmayr2001, Uchiyama1997}. This connection between the violation of a Bell inequality and an internal symmetry of a particle is quite remarkable and must have a deeper and more general meaning.

\emph{Case} b)
When fixing the flavour of the kaons and varying the time of the measurements, it turned out that due to the fast decay compared to the slow oscillation, which increased the mixedness of the total system, a Bell inequality was not violated anymore by QM. However, Beatrix Hiesmayr and a group of experimentalists~\cite{Hiesmayr-DiDomenico-Catalina-etal2012} succeeded to establish a generalized Bell inequality for the $K^0 \bar{K}^0$ system, which was violated by QM in certain measurable time regions. In this case the hidden variable theories were excluded. The experimental preparations for the KLOE-2 detector at DA$\Phi$NE are in progress.\\

I also want to draw attention to possible experiments that test Bell inequalities by inserting a regenerator, that is a piece of matter,  into the kaon beam~\cite{Bramon-Escribano-Garbarino2006, Bramon-Escribano-Garbarino2007, Bramon-Garbarino-PRL88-2002, Bramon-Garbarino-PRL89-2002, Hatice2009}. These experiments are of particular interest since regeneration, a typical quantum feature of the $K$ meson, is directly related to a Bell inequality.

Furthermore, a Bell test for quite a different system, the $\Lambda \bar{\Lambda}$ system, is in preparation~\cite{Hiesmayr-priv-comm2014} and experimentally planned by the FLAIR collaboration, Darmstadt.

Last but not least, tests of local realism in the decay of a charmed particle into entangled vector mesons should be mentioned as well~\cite{Li-Qiao2010, Ding-Li-Qiao2007, Li-Qiao2009}.\\

Considering the quantum correlations of a particle--antiparticle system, the loss of entanglement could be detected via a decoherence process of the system with its environment. Quantitatively, the value of the decoherence parameter corresponded precisely to the amount of entanglement loss~\cite{Bertlmann-Durstberger-Hiesmayr2003}. A number of experiments had been already analyzed~\cite{CPLEAR1998, Bertlmann-Grimus-Hiesmayr1999, Bertlmann-Grimus1998, Bertlmann-Grimus2001}, showing that the produced systems, $K^0 \bar{K}^0$ and $B^0 \bar{B}^0\,$, were indeed entangled over macroscopic distances in accordance with QM. The local realistic theories totally failed to explain the data (for details, see Refs.~\cite{Bertlmann-LectureNotes2006, CarlaSchuler2014, Samitz2012}).\\

These direct Bell-type tests of our basic concepts about matter are of utmost importance since there is always a slim chance of an unexpected result, despite the fact of the general success of quantum mechanics.

There is already a prospering collaboration between theoretical and experimental groups that are planning and performing the tests. The outcome of these experiments will certainly provide extremely valuable knowledge about the nonlocal nature of our matter, in particular, in connection with the inherent symmetries.\\

I think John Bell, who was both, a particle- and a quantum-physicist, had been pleased seeing the developments of this kind of experiments.

\section{Epilogue}

\begin{figure}
\centering
\includegraphics[width=0.4\textwidth]{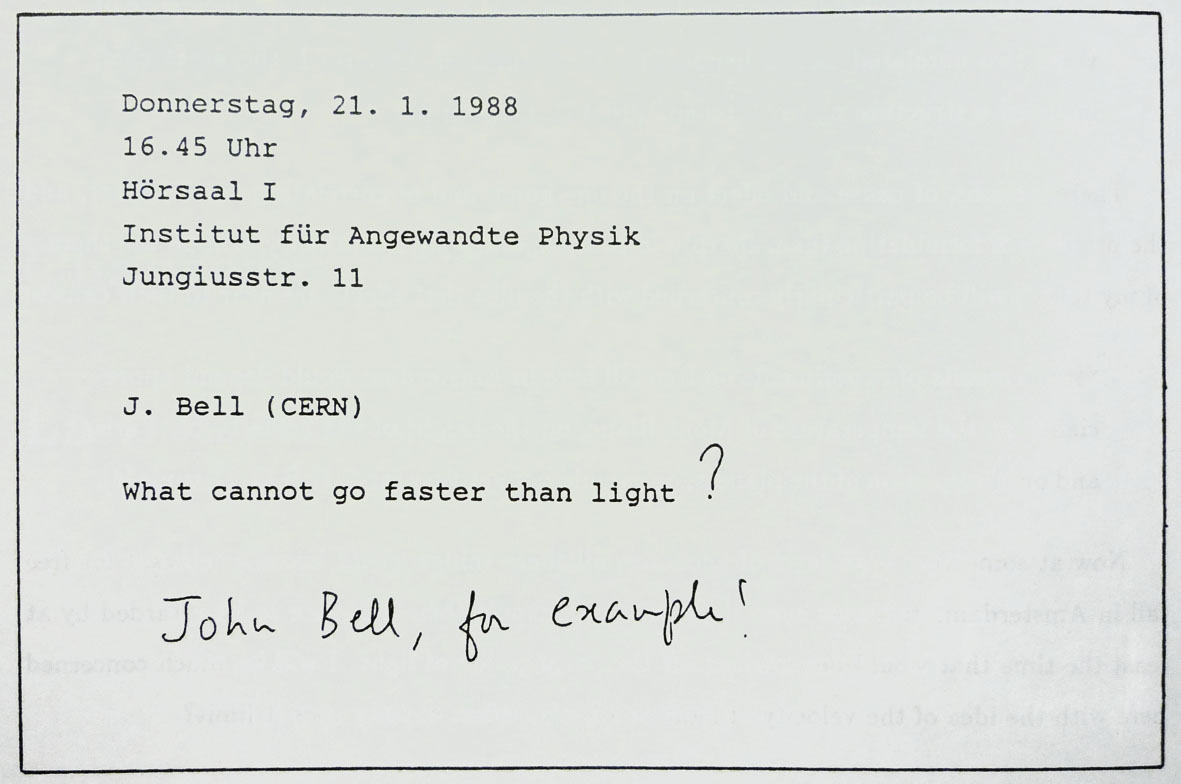}
\caption{Announcement of John Bell's talk: \emph{``What cannot go faster than light?''} at the University of Hamburg in 1988. Somebody with Hanseatic humor added to the announcement by hand: \emph{``John Bell, for example!''}}
	\label{fig:Bell-Vortragsankuendigung}
\end{figure}

When I think again about entanglement, nonlocality and contextuality I can roughly state the following: Entanglement is the appropriate concept for the mathematical and geometric structure of the quantum states but for the physics, the experiments, the concept of nonlocality and contextuality is more adapted.\\

Interestingly, John Bell was not so much concerned about contextuality and its implications. Whereas, I think that contextuality as a quantum feature has a deep rooting in Nature, it is very interesting and still needs to be explored. I am convinced, some day it will have technical applications. Quantum mechanics tells us that the Hilbert spaces of the degrees of freedom of a system are essential. Therefore, it is possible to have entanglement of an internal- with an external-degree of freedom of one single particle, which is, in my opinion, quite remarkable. In this connection, I appreciate very much the neutron interferometer experiments \cite{Hasegawa-etal-KochenSpecker2003, Hasegawa-etal-KochenSpecker2010, Hasegawa-Sponar-etal-BerryPhase2010, Hasegawa-etal-contextuality2009, Bertlmann-Durstberger-Hasegawa-Hiesmayr2004}, where by manipulating the spin (internal degree) one can influence the path (external degree) of the neutron.

The nonlocality feature disturbed John deeply since for him it was equivalent to a \emph{``breaking of Lorentz invariance''} in an extended theory for quantum mechanics, what he hardly could accept. He often remarked:
\emph{``It's a great puzzle to me ... behind the scenes something is going faster than the speed of light."}\\

\begin{figure}
\centering
\includegraphics[width=0.4\textwidth]{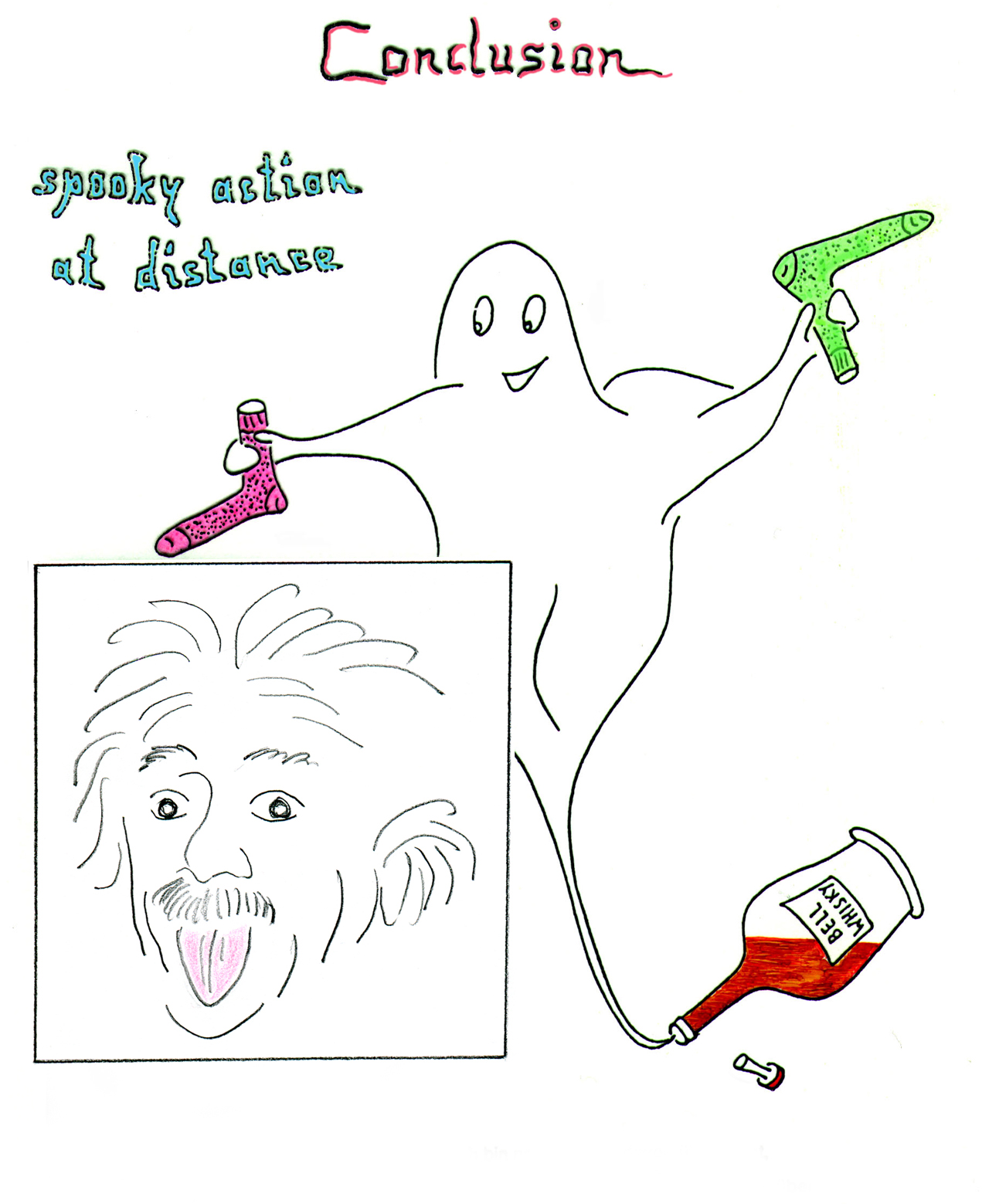}
\caption{Sketch of my conclusions in the paper \emph{``Bell's theorem and the nature of reality''}, which I dedicated John Bell in 1988 on occasion of his 60th birthday~\cite{Bertlmann-BellsTheorem-FoundationOfPhysics}.}
	\label{fig:spooky-ghost}
\end{figure}

 At the end of his Bertlmann's socks paper John expressed again his concern~\cite{Bell-Bsocks-Journal-de-Physique}:

\emph{``It may be that we have to admit that causal influences \textsl{do} go faster than light. The role of Lorentz invariance of a completed theory would then be very problematic. An `aether' would be the cheapest solution. But the unobservability of this aether would be disturbing. So would be the impossibility of `messages' faster than light, which follows from ordinary relativistic quantum mechanics in so far as it is unambiguous  and adequate  for procedures we can actually perform. The exact elucidation of concepts like `message' and `we', would be a formidable challenge.''}\\

In John's last paper \emph{``La nouvelle cuisine''}, published in 1990 \cite{Bell-nouvelle-cuisine} (and see his collected quantum works \cite{Bell-book}), he still remained profoundly concerned with this nonlocal structure of Nature. The paper was based on a talk he gave at the University of Hamburg, in 1988, about the topic: \emph{``What cannot go faster than light?''}. Somebody with Hanseatic humor added to the announcement by hand: \emph{``John Bell, for example!''} (see Fig.~\ref{fig:Bell-Vortragsankuendigung}). This remark made John thinking, what exactly that meant, his whole body or just his legs, his cells or molecules, atoms, electrons ... Was it meant that none of his electrons go faster than light?

In our modern view of Nature the concepts of a classical theory changed, the sharp location of objects had been dissolved by the fuzzyness of the wave function or by the fluctuations in quantum field theory. As John remarked:
\emph{``The concept `velocity of an electron' is now unproblematic only when not thought about it.''}\\

Finally, John discussed \emph{``Cause and effect''} in this paper. As Einstein \cite{Einstein1907} already pointed out, if an effect follows its cause faster than the propagation of light, then there exists an inertial frame where the effect happens before the cause. Such a thing was unacceptable for both, Einstein and Bell. Therefore, sticking to \emph{``no signals faster than light''} Bell defined locally causal theories and demonstrated, via an EPR-Bell type experiment, that \emph{``ordinary quantum mechanics is not locally causal''}, or more precisely, \emph{``quantum mechanics cannot be embedded into a locally causal theory''}. It was essential in his argumentation that the measurement settings $\vec{a}$ at Alice's side and $\vec{b}$ at Bob's side could be chosen totally free, i.e., at random. \emph{``But still, we cannot signal faster than light''} John noted at the end.\\

My personal feeling is that Bell's Theorem, which reveals an apparent nonlocality in Nature, points to a more radical conception whose onset we even do not have yet. It is quite remarkable that this nonlocality cannot be used to send signals with velocity faster than light. Somehow there is a tacit agreement between quantum mechanics and special relativity not to interfere each other. More over, quantum mechanics also does not seem to interfere with general relativity, the contemporary theory of space-time. A consistent quantization of space-time is still lacking, the difficulties are conceptual ones, at the Planck scale the quantum fluctuations of the space-time metric become larger than the considered lengths, so that the metric structure is no longer well-defined. My strong suspicion is, if we can overcome these difficulties once in a \emph{radically} modified conception, this tense status between quantum mechanics and special relativity and the space-time structure is resolved and finds a natural explanation.\\

In a revenge paper: \emph{``Bell's theorem and the nature of reality''}~\cite{Bertlmann-BellsTheorem-FoundationOfPhysics}, which I dedicated John on occasion of his 60th birthday in 1988, I sketched my conclusions in a special figure (see Fig.~\ref{fig:spooky-ghost}). John, as a strict teetotaler, was amused very much by this drawing since the \emph{spooky, nonlocal ghost} emerges from a \emph{Bell Whisky} bottle that \emph{really} does exist.

\begin{acknowledgments}
I would like to thank Renate Bertlmann, Mary Bell, Gregor Weihs and Anton Zeilinger for providing me with photos and Nicolai Friis for illustrating Fig~\ref{fig:Bell-setup-experiment}. I am indebted to Simon Kochen and Eva Ruhnau for drawing attention to reference \cite{GreteHermann1935}, and last but not least, I am thankful to Marissa Giustina for improving my language.
\end{acknowledgments}



\section*{References}


\begin{thebibliography}{10}

\bibitem{Sakurai1973}
J. J. Sakurai, Phys. Lett. B \textbf{46}, 207 (1973)

\bibitem{BellBertlmannDual1980}
J.S. Bell and R.A. Bertlmann, Z. Phys. C \textbf{4}, 11 (1980)

\bibitem{Bertlmann-ActaPhysAustr1981}
R.A. Bertlmann, Acta Phys. Austr. \textbf{53}, 305 (1981)

\bibitem{BellBertlmannMagic1981}
J.S. Bell and R.A. Bertlmann, Nucl. Phys. B \textbf{177}, 218 (1981)

\bibitem{SVZ1979}
M.A. Shifman, V.I. Vainshtein and V.I. Zakharov, Nucl. Phys. B \textbf{147}, 385, 448, 519 (1979)

\bibitem{BertlmannCharmonium-PhysLett1981}
R.A. Bertlmann, Phys. Lett. B \textbf{106}, 336 (1981)

\bibitem{BertlmannCharmonium-NuclPhys1982}
R.A. Bertlmann, Nucl. Phys. B \textbf{204}, 387 (1982)

\bibitem{BellBertlmann-PhysLett1984}
J.S. Bell and R.A. Bertlmann, Phys. Lett. B \textbf{137}, 107 (1984)

\bibitem{BellBertlmann-SVZ1981}
J.S. Bell and R.A. Bertlmann, Nucl. Phys. B \textbf{187}, 285 (1981)

\bibitem{BellBertlmann-LV1983}
R.A. Bertlmann and J.S. Bell, Nucl. Phys. B \textbf{227}, 435 (1983)

\bibitem{QuiggRosner1979}
C. Quigg and J.L. Rosner, Phys. Reports \textbf{56}, 167 (1979)

\bibitem{KrammerKrasemann1979}
M. Krammer and H. Krasemann, Acta Phys. Austr. Suppl. \textbf{21}, 259 (1979)

\bibitem{GrosseMartin1980}
H. Grosse and A. Martin, Phys. Reports \textbf{60}, 341 (1980)

\bibitem{LuchaSchoeberlGromes1991}
W. Lucha, F.F. Sch\" oberl and D. Gromes, Phys. Reports \textbf{200}, 127 (1991)

\bibitem{Bertlmann-GGGG1984}
R.A. Bertlmann, Phys. Lett. B \textbf{148}, 177 (1984)

\bibitem{Leutwyler1981}
H. Leutwyler, Phys. Lett. B \textbf{98}, 447 (1981)

\bibitem{Voloshin1979}
M.B. Voloshin, Nucl. Phys. B \textbf{154}, 365 (1979)

\bibitem{BertlmannPotential1991}
R.A. Bertlmann, Nucl. Phys. B (Proc. Suppl.) \textbf{23}, 307 (1991)

\bibitem{Bell-Bsocks-CERNpreprint}
J.S. Bell, \emph{Bertlmann's socks and the nature of reality}, CERN preprint Ref.TH.2926-CERN, 18 July 1980

\bibitem{Bell-Bsocks-Journal-de-Physique}
J.S. Bell, \emph{Bertlmann's socks and the nature of reality}, Journ. de Phys., Colloque C2, Suppl. 3, Tome 42, p. C2-41, 1981

\bibitem{EPR}
A. Einstein, B. Podolsky, and N.Rosen, Phys. Rev. \textbf{47}, 777 (1935)

\bibitem{Aspect-Dalibard-Roger1982}
A. Aspect, J. Dalibard, and G. Roger, Phys. Rev. Lett. \textbf{49}, 1804 (1982)

\bibitem{Mermin-RevModPhys}
N.D. Mermin, Rev. Mod. Phys. \textbf{65}, 803 (1993)

\bibitem{Bell-RevModPhys1966}
J.S. Bell, Rev. Mod. Phys. \textbf{38}, 447 (1966)

\bibitem{vonNeumann1932}
J. von Neumann, \emph{Mathematische Grundlagen der Quantenmechanik}, Springer, Berlin 1932; English translation: Princeton University Press 1955

\bibitem{GreteHermann1935}
G. Hermann, \emph{Die naturphilosophischen Grundlagen der Quantenmechanik}, Abhandlungen der Fries'schen Schule, Band \textbf{6}, 69, (1935)

\bibitem{Kochen2002}
S.B. Kochen,  \emph{The geometry of quantum paradoxes}, in \emph{Quantum [Un]speakables}, R.A. Bertlmann and A. Zeilinger (eds.), p. 257, Springer 2002

\bibitem{Gleason1957}
A.M. Gleason, J. Math. Mech. \textbf{6}, 885 (1957)

\bibitem{Kochen-Specker}
S.B. Kochen and E. Specker, J. Math. Mech. \textbf{17}, 59 (1967)

\bibitem{Cabello-Guehne-etal_PRL2013}
E. Amselem, M. Bourennane, C. Budroni, A. Cabello, O. G\"uhne, M. Kleinmann, J.-\AA. Larsson, and M. Wie\'sniak, Phys. Rev. Lett. \textbf{110}, 078901 (2013)

\bibitem{Canas-Cabello-etal2013}
G. Ca\~nas, S. Etcheverry, E.S. G\'omez, C. Saavedra, G.B. Xavier, G. Lima, and A. Cabello, Phys. Rev. A \textbf{90}, 012119 (2014)

\bibitem{Cabello-etal_PRL-112-040401-2014}
A. Cabello, S. Severini, and A. Winter, Phys. Rev. Lett. \textbf{112}, 040401 (2014)

\bibitem{Cabello-etal_PRL-111-180404-2013}
A. Cabello, P. Badziag, M. Terra Cunha, and M. Bourennane, Phys. Rev. Lett. \textbf{111}, 180404 (2013)

\bibitem{Guehne-Cabello-etal2013}
O. G\" uhne, C. Budroni, A. Cabello, M. Kleinmann, and J.-\r A. Larsson, Phys. Rev. A \textbf{89}, 062107 (2014)

\bibitem{Acin-etal2012}
A. Ac\'in, T. Fritz, A. Leverrier, and A. Bel\'en Sainz, \emph{A combinatorial approach to nonlocality and contextuality}, arXiv:1212.4084 (2012)

\bibitem{Bohm1952}
D. Bohm, Phys. Rev. \textbf{85}, 166 (1952); ibid p. 180 (1952)

\bibitem{Einstein-Born-letter}
A. Einstein, Letter of to M. Born, May 12th, 1952, in \emph{Albert Einstein Max Born, Briefwechsel 1916 - 1955}, Rowohlt 1969

\bibitem{Bohm-Aharanov-EPRspin1957}
D. Bohm and Y. Aharonov, Phys. Rev. \textbf{108}, 1070 (1957)

\bibitem{Bell-Physics1964}
J.S. Bell, Physics \textbf{1}, 195 (1964)

\bibitem{CHSH}
J.F. Clauser, M.A. Horne, A. Shimony, R.A. Holt, Phys. Rev. Lett. \textbf{23}, 880 (1969)

\bibitem{Bell-Fermi-School-1971}
J.S. Bell, \emph{Introduction to the hidden variable question}, in \emph{Foundations of Quantum Mechanics}, Proceedings of the International School of Physics ``Enrico Fermi'', pp 171-181, Academic New York 1971

\bibitem{nielsen-chuang2000}
M.A. Nielsen and I.L. Chuang, \emph{Quantum Computation and Quantum Information}, Cambridge 2000

\bibitem{bertlmann-zeilinger02}
R.A. Bertlmann and A. Zeilinger (eds.), \emph{Quantum [Un]speakables}, Springer 2002

\bibitem{Wigner1970}
E.P. Wigner, Am. J. Phys. \textbf{38}, 1005 (1970)

\bibitem{Clauser-Horne1974}
J.F. Clauser and M.A. Horne, Phys. Rev. D \textbf{10}, 526 (1974)

\bibitem{Brunner-etal-Bell-nonlocality2013}
N. Brunner, D. Cavalcanti, S. Pironio, V. Scarani, and S. Wehner, Rev. Mod. Phys. \textbf{86}, 419 (2014)

\bibitem{werner89}
R.F. Werner, Phys. Rev. A \textbf{40}, 4277 (1989)

\bibitem{Gisin1991}
N. Gisin, Phys. Lett. A \textbf{154}, 201 (1991)

\bibitem{Barrett2002}
J. Barrett, Phys. Rev. A \textbf{65}, 042302 (2002)

\bibitem{Acin-Gisin-Toner2006}
A. Ac\'\i n, N. Gisin and B. Toner, Phys. Rev. A \textbf{73}, 062105 (2006)

\bibitem{Vertesi2008}
T. V\'ertesi, Phys. Rev. A \textbf{78}, 032112 (2008)

\bibitem{peres96}
A. Peres, Phys. Rev. Lett. \textbf{77}, 1413 (1996)

\bibitem{horodecki96}
M. Horodecki, P. Horodecki, and R. Horodecki, Phys. Lett. A \textbf{223}, 1 (1996)

\bibitem{horodecki98}
P. Horodecki, M. Horodecki, and R. Horodecki, Phys. Rev. Lett. \textbf{80}, 5239 (1998)

\bibitem{HHHH07}
R. Horodecki, P. Horodecki, M. Horodecki, and K. Horodecki, Rev. Mod. Phys. \textbf{81}, 865 (2009)

\bibitem{bertlmann-krammer-AnnPhys09}
R.A. Bertlmann and P. Krammer, Ann. Phys. \textbf{324}, 1388 (2009)

\bibitem{bertlmann-krammer-PRA-78-08}
R.A. Bertlmann and P. Krammer, Phys. Rev. A \textbf{78}, 014303 (2008)

\bibitem{bertlmann-krammer-PRA-77-08}
R.A. Bertlmann and P. Krammer, Phys. Rev. A \textbf{77}, 024303 (2008)

\bibitem{baumgartner-hiesmayr-narnhofer06}
B. Baumgartner, B.C. Hiesmayr, and H. Narnhofer, Phys. Rev. A \textbf{74}, 032327 (2006)

\bibitem{baumgartner-hiesmayr-narnhofer07}
B. Baumgartner, B.C. Hiesmayr, and H. Narnhofer, J. Phys. A: Math. Theor. \textbf{40}, 7919 (2007)

\bibitem{baumgartner-hiesmayr-narnhofer08}
B. Baumgartner, B.C. Hiesmayr, and H. Narnhofer, Phys. Lett. A \textbf{372}, 2190 (2008)

\bibitem{horodecki95}
R. Horodecki, P. Horodecki, and M. Horodecki, Phys. Lett. A \textbf{200}, 340 (1995)

\bibitem{terhal00}
B.M. Terhal, Phys. Lett. A \textbf{271}, 319 (2000)

\bibitem{bertlmann-narnhofer-thirring02}
R.A. Bertlmann, H. Narnhofer, and W. Thirring, Phys. Rev. A \textbf{66}, 032319 (2002)

\bibitem{bruss02}
D. Bru\ss, J. Math. Phys. \textbf{43}, 4237 (2002)

\bibitem{bruss-cirac-horodecki-etal02}
D. Bru\ss, J.I. Cirac, P. Horodecki, F. Hulpke, B. Kraus, M. Lewenstein, and A. Sanpera, J. Mod. Opt. \textbf{49}, 1399 (2002)

\bibitem{guehne-toth-PhysRep2009}
O. G\" uhne and G. T\' oth, Phys. Rep. \textbf{474}, 1 (2009)

\bibitem{krammer-DA-2005}
P. Krammer, \emph{Quantum entanglement: Detection, classification, and quantification}, Diploma Thesis, University of Vienna, 2005

\bibitem{gabriel-DA-2009}
A. Gabriel, \emph{Quantum entanglement and geometry}, Diploma Thesis, University of Vienna, 2009

\bibitem{vollbrecht-werner-PRA00}
K.G.H. Vollbrecht and R.F. Werner, Phys. Rev. A \textbf{64}, 062307 (2000)

\bibitem{horodecki-R-M96}
R. Horodecki and M. Horodecki, Phys. Rev. A \textbf{54}, 1838 (1996)

\bibitem{TBKN}
W. Thirring, R.A. Bertlmann, P. K\" ohler, and H. Narnhofer, Eur. Phys. J. D \textbf{64}, 181 (2011)

\bibitem{kus-zyczkowski}
M. Ku\'s and K. \.Zyczkowski, Phys. Rev. A \textbf{63}, 032307 (2001)

\bibitem{zyczkowski-bengtsson}
K. \.Zyczkowski and I. Bengtsson, arXiv:quant-ph/0606228

\bibitem{bengtsson-zyczkowski-book}
I. Bengtsson and K. \.Zyczkowski, \emph{Geometry of Quantum states: An Introduction to Quantum Entanglement}, Cambridge University Press 2006

\bibitem{Bohr}
N. Bohr, Phys. Rev. \textbf{48}, 696 (1935)

\bibitem{deBroglie1960}
L. de Broglie, \emph{Nonlinear wave mechanics: A causal interpretation}, Elsevier, Amsterdam 1960

\bibitem{deBroglie1963}
L. de Broglie, \emph{Introduction to the Vigier theory of elementary particles}, Elsevier, Amsterdam 1963

\bibitem{deBroglie1964}
L. de Broglie, \emph{The current interpretation of wave mechanics: A critical study}, Elsevier, Amsterdam 1964

\bibitem{Pauli-letter-to-Born1954}
Letter of W. Pauli to M. Born, March 31st, 1954, in \emph{Wolfgang Pauli, scientific correspondence with Bohr, Einstein, Heisenberg}, Vol. IV, Part II, 1953-1954, ed. Karl von Meyenn, Springer 1999

\bibitem{Clauser2002}
J.F. Clauser, \emph{Early history of Bell's theorem}, in \emph{Quantum [Un]speakables}, R.A. Bertlmann and A. Zeilinger (eds.), p. 61, Springer 2002

\bibitem{Clauser1969}
J.F. Clauser, Bull. Am. Phys. Soc. \textbf{14}, 578 (1969)

\bibitem{Clauser-Freedman1972}
S.J. Freedman and J.F. Clauser, Phys. Rev. Lett. \textbf{28}, 938 (1972)

\bibitem{Freedman1972}
S.J. Freedman, LBL Rep. No. \textbf{191}, Lawrence Berkeley National Laboratory (1972)

\bibitem{Holt-Pipkin1974}
R.A. Holt and F.M. Pipkin, \emph{Quantum mechanics versus hidden variables: polarization correlation measurement on an atomic mercury cascade}, Harvard preprint, 1974

\bibitem{Clauser1976}
J.F. Clauser, Phys. Rev. Lett. \textbf{36}, 1223 (1976)

\bibitem{Fry-Thompson1976}
E.S. Fry and R.C. Thompson, Phys. Rev. Lett. \textbf{37}, 465 (1976)

\bibitem{Kasday-etal1970}
L.R. Kasday, J.D. Ullmann, and C.S. Wu, Bull. Am. Phys. Soc. \textbf{15}, 586 (1970); Nouvo Cim. B \textbf{25}, 633 (1975)

\bibitem{Faraci-etal1974}
G. Faraci, D. Gutkowski, S. Notarrigo, and A.R. Pennisi, Nouvo Cim. Lett. \textbf{9}, 607 (1974)

\bibitem{Wilson-etal1976}
A.R. Wilson, J. Lowe, and D.K. Butt, J. Phys. G \textbf{2}, 613 (1976)

\bibitem{Bruno-etal1977}
M. Bruno, M. D'Agostino, and C. Maroni, Nouvo Cim. B \textbf{40}, 143 (1977)

\bibitem{Bertolini-etal1981}
G. Bertolini, E. Diana, and A. Scotti, Nouvo Cim. B \textbf{63}, 651 (1981)

\bibitem{Lamehi-Rachti-Mittig1976}
M. Lamehi-Rachti and W. Mittig, Phys. Rev. D \textbf{14}, 2543 (1976)

\bibitem{Aspect1976}
A. Aspect, Phys. Rev. D \textbf{14}, 1944 (1976)

\bibitem{Aspect-etal1981}
A. Aspect, P. Grangier, and G. Roger, Phys. Rev. Lett. \textbf{47}, 460 (1981)

\bibitem{Aspect-etal1982}
A. Aspect, P. Grangier, and G. Roger, Phys. Rev. Lett. \textbf{49}, 91 (1982)

\bibitem{Aspect-etal1985}
A. Aspect and P. Grangier, Lett. Nuovo Cim. \textbf{43}, 345 (1985)

\bibitem{Kleinpoppen-etal1985}
W. Perrie, A.J. Duncan, H.J. Beyer, and H. Kleinpoppen, Phys. Rev. Lett. \textbf{54}, 1790 (1985); ibid \textbf{54}, 2647 (1985)

\bibitem{Weihs-Zeilinger-Bell-experiment1998}
G. Weihs, T. Jennewein, C. Simon, H. Weinfurter, and A. Zeilinger, Phys. Rev. Lett. \textbf{81}, 5039 (1998)

\bibitem{Weihs-thesis-Bell-experiment1998}
G. Weihs, \emph{Ein Experiment zum Test der Bellschen Ungleichung unter Einstein Lokalit\" at}, PhD Thesis, University of Innsbruck, 1998

\bibitem{Brendel-etal1992}
J. Brendel, E. Mohler, and W. Martienssen, Eur. Phys. Lett. \textbf{20}, 575 (1992)

\bibitem{Tapster-etal1994}
P.R. Tapster, J.G. Rarity, and P.C.M. Owens, Phys. Rev. Lett. \textbf{73}, 1923 (1994)

\bibitem{Tittel-Gisin-etal1998}
W. Tittel, J. Brendel, H. Zbinden, and N. Gisin, Phys. Rev. Lett. \textbf{81}, 3563 (1998)

\bibitem{Bouwmeester-Zeilinger-etal1997}
D. Bouwmeester, J.-W. Pan, K. Mattle, M. Eibl, H. Weinfurter, and A. Zeilinger, Nature \textbf{390}, 575 (1997)

\bibitem{Ma-Zeilinger-etal-teleportation-144km-Nature2012}
X.-S. Ma, T. Herbst, T. Scheidl, D. Wang, S. Kropatschek, W. Naylor, B. Wittmann, A. Mech, J. Kofler, E. Anisimova, V. Makarov,	T. Jennewein, R. Ursin, and A. Zeilinger, Nature \textbf{489}, 269 (2012)

\bibitem{Jennewein-Zeilinger-etal-QuCrypto2000}
T. Jennewein, C. Simon, G. Weihs, H. Weinfurter, and A. Zeilinger, Phys. Rev. Lett. \textbf{84}, 4729 (2000)

\bibitem{Gisin-etal-QuCrypto2002}
N. Gisin, G. Ribordy, W. Tittel, and H. Zbinden, Rev. Mod. Phys. \textbf{74}, 145 (2002)

\bibitem{Resch-Blauensteiner-Zeilinger-etal-freespace-City2005}
K.J. Resch, M. Lindenthal, B. Blauensteiner, H.J. B\" ohm, A. Fedrizzi, C. Kurtsiefer, A. Poppe, T. Schmitt-Manderbach, M. Taraba, R. Ursin, P. Walther, H. Weier, H. Weinfurter, and A. Zeilinger, Optics Express \textbf{13}, 202 (2005)

\bibitem{Ursin-Blauensteiner-Zeilinger-etal-Tenerife2007}
R. Ursin, F. Tiefenbacher, T. Schmitt-Manderbach, H. Weier, T. Scheidl, M. Lindenthal, B. Blauensteiner, T. Jennewein, J. Perdigues, P. Trojek, B. \" Omer, M. F\" urst, M. Meyenburg, J. Rarity, Z. Sodnik, C. Barbieri, H. Weinfurter, and A. Zeilinger, Nature Physics \textbf{3}, 481 (2007)

\bibitem{Giustina-Zeilinger-fairsampling-Nature2013}
M. Giustina, A. Mech, S. Ramelow, B. Wittmann, J. Kofler, J. Beyer, A. Lita, B. Calkins, T. Gerrits, S.W. Nam, R. Ursin, and A. Zeilinger, Nature \textbf{497}, 227 (2013)

\bibitem{Eberhard-inequ-1993}
P.H. Eberhard, Phys. Rev. A \textbf{47}, 747 (1993)

\bibitem{Rowe-Wineland-Bell-ions-Nature2001}
M.A. Rowe, D. Kielpinsky, V. Meyer, C.A. Sackett, W.M. Itano, and D.J. Wineland, Nature \textbf{409}, 791 (2001)

\bibitem{Hasegawa-etal-KochenSpecker2003}
Y. Hasegawa, R. Loidl, G. Badurek, M. Baron, and H. Rauch, Nature \textbf{425}, 45 (2003)

\bibitem{Hasegawa-etal-KochenSpecker2010}
Y. Hasegawa, K. Durstberger-Rennhofer, S. Sponar, and H. Rauch, \emph{Kochen-Specker theorem studied with neutron interferometer}, arXiv:1004.2836 (2010)

\bibitem{Hasegawa-etal-contextuality2009}
H. Bartosik, J. Klepp, C. Schmitzer, S. Sponar, A. Cabello, H. Rauch, and Y. Hasegawa, Phys. Rev. Lett. \textbf{103}, 040403 (2009)

\bibitem{Bertlmann-Hiesmayr2001}
R.A. Bertlmann and B.C. Hiesmayr, Phys. Rev. A \textbf{63}, 062112 (2001)

\bibitem{Bertlmann-LectureNotes2006}
R.A. Bertlmann, \emph{Entanglement, Bell inequalities and decoherence in particle physics}, in \emph{Quantum coherence: From quarks to solids}, eds. W. P\" otz, J. Fabian, and U. Hohensteiner, Lect. Notes Phys. \textbf{689}, 1-45, Springer 2006

\bibitem{CarlaSchuler2014}
C. Schuler, \emph{Entanglement, Bell inequalities and decoherence in neutral K-meson systems}, Bachelor Thesis, University of Vienna, 2014

\bibitem{Bertlmann-Bramon-Garbarino-Hiesmayr2004}
R.A. Bertlmann, A. Bramon, G. Garbarino, and B.C. Hiesmayr, Phys. Lett. A \textbf{332}, 355 (2004)

\bibitem{Amelino-Camelia-KLOEcoll2010}
G. Amelino-Camelia et al., KLOE collaboration, Eur. Phys. J. C \textbf{68}, 619 (2010)

\bibitem{Bertlmann-Grimus-Hiesmayr2001}
R.A. Bertlmann, W. Grimus, and B.C. Hiesmayr, Phys. Lett. A \textbf{289}, 21 (2001)

\bibitem{Uchiyama1997}
F. Uchiyama, Phys. Lett. A \textbf{231}, 295 (1997)

\bibitem{Hiesmayr-DiDomenico-Catalina-etal2012}
B.C. Hiesmayr, A. Di Domenico, C. Curceanu, A. Gabriel, M. Huber, J-A. Larsson, and P. Moskal, Eur. Phys. J. C \textbf{72}, 1856 (2012)

\bibitem{Bramon-Escribano-Garbarino2006}
A. Bramon, R. Escribano, and G. Garbarino, Found. Phys. \textbf{36}, 563 (2006)

\bibitem{Bramon-Escribano-Garbarino2007}
A. Bramon, R. Escribano, and G. Garbarino, \emph{A review on Bell inequality tests with neutral kaons}, Frascati Phys. Ser., Vol. XLIII, ed. A. Di Domenico, p. 217, 2007

\bibitem{Bramon-Garbarino-PRL88-2002}
A. Bramon and G. Garbarino, Phys. Rev. Lett. \textbf{88}, 040403 (2002)

\bibitem{Bramon-Garbarino-PRL89-2002}
A. Bramon and G. Garbarino, Phys. Rev. Lett. \textbf{89}, 160401 (2002)

\bibitem{Hatice2009}
H. Tataroglu, \emph{Nichtlokale Korrelation in Kaonischen Systemen}, Diploma Thesis, University of Vienna, 2009

\bibitem{Hiesmayr-priv-comm2014}
B.C. Hiesmayr, privat communication, 2014

\bibitem{Li-Qiao2010}
J.-L. Li and C.-F. Qiao, Sci. China G \textbf{53}, 870 (2010)

\bibitem{Ding-Li-Qiao2007}
Y.-B. Ding, J.-L. Li, and C.-F. Qiao, High En. Phys. and Nucl. Phys. \textbf{31}, 1086 (2007)

\bibitem{Li-Qiao2009}
J.-L. Li and C.-F. Phys. Lett. A \textbf{373}, 4311 (2009)

\bibitem{Bertlmann-Durstberger-Hiesmayr2003}
R.A. Bertlmann, K. Durstberger, and B. C. Hiesmayr, Phys.Rev. A \textbf{68}, 012111 (2003)

\bibitem{CPLEAR1998}
A. Apostolakis et al: CPLEAR Coll., Phys. Lett. B \textbf{422}, 339 (1998)

\bibitem{Bertlmann-Grimus-Hiesmayr1999}
R.A. Bertlmann, W. Grimus, and B. C. Hiesmayr, Phys.Rev. D \textbf{60}, 114032 (1999)

\bibitem{Bertlmann-Grimus1998}
R.A. Bertlmann and W. Grimus, Phys.Rev. D \textbf{58}, 034014 (1998)

\bibitem{Bertlmann-Grimus2001}
R.A. Bertlmann and W. Grimus, Phys.Rev. D \textbf{64}, 056004 (2001)

\bibitem{Samitz2012}
D. Samitz, \emph{Particle oscillations, entanglement and decoherence}, Bachelor Thesis, University of Vienna, 2012

\bibitem{Hasegawa-Sponar-etal-BerryPhase2010}
S. Sponar, J. Klepp, K. Durstberger-Rennhofer, R. Loidl, S. Filipp, M. Lettner, R. A. Bertlmann, G. Badurek, H.Rauch, and Y. Hasegawa, J. Phys. A: Math. Theor. 43, 354015 (2010)

\bibitem{Bertlmann-Durstberger-Hasegawa-Hiesmayr2004}
R.A. Bertlmann, K. Durstberger, Y. Hasegawa, and B.C. Hiesmayr, Phys. Rev. A \textbf{69}, 032112 (2004)

\bibitem{Bell-nouvelle-cuisine}
J.S. Bell, \emph{La nouvelle cuisine}, in \emph{Between Science and Technology}, eds. A. Sarlemijn and P. Kroes, Elsevier Science Publishers 1990

\bibitem{Bell-book}
J.S. Bell, \emph{Speakable and unspeakable in quantum mechanics}, Cambridge University Press, Second Ed., 2004

\bibitem{Einstein1907}
A. Einstein, Ann. Phys. \textbf{23}, 371 (1907)

\bibitem{Bertlmann-BellsTheorem-FoundationOfPhysics}
R.A. Bertlmann, \emph{``Bell's theorem and the nature of reality''}, preprint University of Vienna, 1988, and Found. Phys. \textbf{20}, 1191 (1990)



\end{thebibliography}
\end{document}